\documentclass[12pt]{article} 
\usepackage[hidelinks]{hyperref}
\usepackage{graphicx,multirow}
\usepackage{amssymb,amsmath,amsfonts,bm}
\usepackage{setspace}
\usepackage{natbib}
\usepackage{multicol}
\usepackage{caption}
\usepackage{xcolor}
\usepackage[misc]{ifsym}
\usepackage{colortbl}
\usepackage{titlesec}

\newcommand{\be}{\begin{equation*}}
\newcommand{\ee}{\end{equation*}}
\newcommand{\bne}{\begin{equation}}
\newcommand{\ene}{\end{equation}}
\newcommand{\co}{\cellcolor{orange!80}}
\newcommand{\cg}{\cellcolor{green}}
\newcommand{\cre}{\cellcolor{red!90}}
\newcommand{\cy}{\cellcolor{yellow}}

\begin{document}

\title{Binomial confidence intervals for rare events: importance of defining margin of error relative to magnitude of proportion}
\author{Owen McGrath \footnote{Corresponding author. University of Limerick, Ireland; owen.mcgrath@ul.ie}  \hspace{3cm}
Kevin Burke\footnote{University of Limerick, Ireland; kevin.burke@ul.ie \newline\newline\indent\indent
The corresponding author acknowledges the financial support of Limerick Institute of Technology.} }
\date{\today}

\maketitle

\begin{abstract}

Confidence interval performance is typically assessed in terms of two criteria: coverage probability and interval width (or margin of error). In this paper, we assess the performance of four common proportion interval estimators: the Wald, Clopper-Pearson (exact), Wilson and Agresti-Coull, in the context of rare-event probabilities. We define the interval precision in terms of a relative margin of error which ensures consistency with the magnitude of the proportion. Thus, confidence interval estimators are assessed in terms of achieving a desired coverage probability whilst simultaneously satisfying the specified relative margin of error. We illustrate the importance of considering both coverage probability and relative margin of error when estimating rare-event proportions, and show that within this framework, all four interval estimators perform somewhat similarly for a given sample size and confidence level. We identify relative margin of error values that result in satisfactory coverage whilst being conservative in terms of sample size requirements, and hence suggest a range of values that can be adopted in practice. The proposed relative margin of error scheme is evaluated analytically, by simulation, and by application to a number of recent studies from the literature.\\


\end{abstract}

\qquad

\newpage


\section{Introduction}
\label{sec:Intro}

A fundamental problem in applied statistics is the construction of a confidence interval (CI) for a binomial proportion, $p$. In many applications, one deals with a large population within which an event of interest is rare. For example, in clinical statistics, $p$ could represent the proportion of patients exhibiting treatment side effects; such a scenario arose in the context of COVID-19 vaccination \citep{Polack:2020}. In manufacturing, the number of defective components is often very small relative to the large number of components produced. Indeed, many manufacturers now achieve a defect rate of 3.4 in one million \citep{Evans:2015,Woodall:2014}. In the aviation industry, strict regulations  ensure that safety incidents are deemed as a rare occurrence, for example, \cite{Boeing} shows that very few incidents occur within a large sample of flights. (Note: we revisit the COVID-19 and aviation examples as case studies in Section \ref{sec:Data}, along with an ADHD medication example.)

Ascertaining the order of magnitude of $p$ is important in ``large populations" such as the aforementioned. Indeed, with population sizes in the millions (or billions), there is a big practical difference between $p=10^{-4}$ and $p=10^{-6}$ (but such differences are much less important/detectable in smaller populations). For example, in high-throughput manufacturing, a difference in the order of magnitude in the failure rate has significant implications for the number of defects and/or product returns. From a purely pragmatic perspective, note that ten thousand observations are needed to obtain, on average, one event when $p=10^{-4}$. However, while it is expected that relatively large samples will be required to adequately estimate the order of magnitude of a small proportion, $p$, practitioners will need more specific guidance on the sample size requirements; this is not well covered by the existing literature.

The problem of constructing a CI for $p$ has a wide literature, including several comparative studies, for example, \cite{Goncalves:2012}, \cite{Leemis:1996}, \cite{Newcombe:1998} and \cite{Pires:2008}. These works assess various proportion estimators, for example, \cite{Pires:2008} compare twenty different methods. However, these works, and the literature in general, focus primarily on situations where $p$ is moderately large. As such, there is much less guidance in the existing literature regarding the scenario where $p$ is small. Furthermore, there is little discussion of relative margin of error, which is needed in this small $p$ setting. Whereas relative margin of error is not a prominent feature of CI assessment for moderately large proportions, it is essential that the margin of error scales with the magnitude of $p$ for rare events. Therefore, we consider a valid CI estimator as one that achieves a desired coverage probability whilst also maintaining a specified relative margin of error, and, in contrast to much of the existing literature, we focus on the small $p$ regime of $p\in [10^{-6}, 10^{-1}]$, where relative margin of error is especially important.

For our analysis, we consider the most widely used binomial confidence interval, the Wald interval, along with three other common intervals: Clopper-Pearson (exact), Wilson (score) and Agresti-Coull (adjusted Wald) \citep{Agresti:1998, Clopper:1934, Wilson:1927}. Despite its widespread use, the Wald interval is known to produce inadequate coverage when $p$ is near 0 or 1, and/or the sample size, $n$, is small. It has also been well documented that this interval suffers from erratic coverage, even when $p$ is moderate \citep{Agresti:1998,Blyth:1983,Bohning:1994,Vollset:1993}. \cite{Brown:2001} show that this coverage fluctuation occurs for large $n$ and recommend against using the Wald interval in practice.  \cite{Newcombe:1998} also discourages the use of the Wald interval and suggests that its use be restricted to sample size planning. In recent work, \cite{Andersson:2023} discusses the deficiencies of the Wald interval and examines its coverage and noncoverage performance relative to the Wilson interval. Whilst the criticisms of the Wald interval can be justified, particularly when $n$ is small, it is worth noting that the issue of erratic coverage is not unique to the Wald interval; this behaviour is related to the binomial distribution and we illustrate (in Section \ref{sec:RelMoE}) that it occurs for all four interval estimators.

A common approach in determining sample sizes is to set the (Wald) CI margin of error equal to a specified value, $\epsilon$, and then solve for $n$. In order to maintain consistency between $\epsilon$ and $p$, we consider the relative margin of error, $\epsilon_R=\epsilon/p$, and obtain sample sizes by setting $\epsilon_R$ to a specified value and solving each interval equation for $n$. \cite{Lwanga:1991} provide (Wald) sample size calculations for fixed and relative margins of error in the range $[0.01, 0.5]$ for $p\in[0.05, 0.95]$. However, in our work, we focus on the small-$p$ regime of $p \in [10^{-6}, 10^{-1}]$ and provide computed coverage probabilities relating to $\epsilon_R \in [0.05,0.75]$. 
In this regime, it is important to consider relative precision over fixed precision (fixed $\epsilon$ value). For example, $\epsilon=0.1$ might be considered as reasonable precision for $p=0.4$, but could
equally be considered reasonable for $p = 0.2$. However, where $\epsilon=0.05$ could be considered as a valid margin of error for $p=10^{-1}$, it is far too large for a success probability of the order $p=10^{-3}$. Ultimately, we find that  $\epsilon_R \in [0.1, 0.5]$ yields a good compromise between estimation precision, coverage performance, and sample size requirements.

In this work, we illustrate the importance of using relative margin of error in a small $p$ regime, recommend practical tolerances for both relative margin of error and coverage probability, and provide comparisons in terms of sample size requirements. We show that when CI performance is assessed in terms of both coverage probability and relative margin of error, the four CI estimators perform similarly in many cases. Although the differences between the estimators is less pronounced when considering the relative margin of error, we show that the Wilson (score) interval provides the best overall performance. In addition, we provide practical guidance on the sample sizes required to attain reasonable CI performance for various (small) values of $p$. We anticipate that this guidance will be useful to researchers working in real-world small-$p$ applications.

The remainder of this article is organised as follows. In Section \ref{sec:Estimators}, we review some of the proportion estimators proposed in the literature, focusing, in particular, on the Wald, Clopper-Pearson, Wilson and Agresti-Coull intervals.  Section \ref{sec:Criteria} provides details of the CI evaluation criteria used in the work. Section \ref{sec:Sample} briefly discusses the initial estimation of $p$ for sample size planning, then moves on to using relative margin of error in such planning, and CI performance evaluation. Sections \ref{sec:RelMoE} and \ref{sec:CITable} illustrate the importance of employing a relative margin of error scheme in CI assessment. Section \ref{sec:Smalln} presents the challenge of estimating a rare-event proportion using a small sample size. In Section \ref{sec:Data}, we present a number of case studies to demonstrate the relative margin of error schemes in assessing the validity of estimated intervals. Finally, the article concludes in Section \ref{sec:Conclusion} with a discussion.


\section{Binomial Proportion Interval Estimators}
\label{sec:Estimators}

Several methods have been devised to estimate a binomial proportion, $p$, including the Wald, Clopper-Pearson, Wilson, Agresti-Coull, Jeffreys, arcsine transformation, Jeffreys' Prior and the likelihood ratio interval. A range of ensemble approaches have also been considered, for example,  \cite{Kabaila:2016}, \cite{Park:2019} and \cite{Turek:2012}. In this work we assess the performance of the Wald, Clopper-Pearson, Wilson and Agresti-Coull intervals in standard form, i.e., without modification or application of continuity correction. 


\subsection{\textit{Details of the Intervals Considered in our Work}}
\label{sec:Est_details}
The Wald interval is included in this study as it is the most widely known and used estimator. We assess the \cite{Clopper:1934} interval as it is an exact method, i.e., unlike the other estimators used in this work, it is based directly on the cumulative probabilities of the binomial distribution. For this reason it is often regarded as the ``gold standard" in binomial proportion estimation \citep{Agresti:1998,Goncalves:2012,Newcombe:1998}. Where the Wald interval is known to produce inadequate coverage for small $n$ or $p$, the Clopper-Pearson interval is generally regarded as being overly conservative, unless $n$ is quite large \citep{Agresti:1998,Brown:2001,Newcombe:1998,Thulin:2014}. The intervals proposed by \cite{Wilson:1927} and \cite{Agresti:1998} both offer a compromise between the (liberal) Wald interval, and the (conservative) Clopper-Pearson interval.

We assess the performance of the Wilson and Agresti-Coull intervals in this work given their popularity within the literature. For example, \cite{Brown:2001} recommend the Wilson or Jeffreys interval for small $n$. For larger $n$, they recommend the Wilson, Jeffreys or Agresti-Coull intervals, preferring the Agresti-Coull method for its simpler presentation. \cite{Agresti:1998} recommend the Wilson interval, and add that the $95\%$ Wilson interval has similar performance to their method. \cite{Newcombe:1998} remarks on the \textit{mid-$p$} method (an exact method closely related to the Clopper-Pearson method), and the Wilson method, noting the Wilson's advantage of having a simple closed form. \cite{Vollset:1993} too recommends the mid-$p$ and Wilson (uncorrected and continuity corrected) intervals, along with the Clopper-Pearson interval, stating that those four intervals can be safely used at all times. \cite{Pires:2008} recommend the continuity corrected arcsine transformation or the Agresti-Coull method. \cite{Krishnamoorthy:2007} show that when controlling for Type I and Type II error rates in one-sided hypotheses, the Wilson (score) and exact (Clopper-Pearson) tests require the same sample size. However, for two-sided hypothesis applications, and for constructing confidence intervals, they recommend the Wilson method.

Listed below are the Wald (W), Clopper-Pearson (CP), Wilson (WS) and Agresti-Coull (AC) interval formulas, where $z_{\alpha/2}$ denotes the $1-\alpha/2$ quantile of the standard normal distribution and $x$ is the number of successes in a sample size of size $n$.
\vspace{-0.25cm}
\begin{align}
\nonumber \text{W:\quad} & \widehat{p}\pm z_{\alpha/2}\sqrt{\dfrac{\widehat{p}(1-\widehat{p})}{n}}, \text{where } \widehat{p} = x/n.\\[0.1cm]
\nonumber \text{CP:\quad} & \text{Lower: Beta}\left({\alpha}/{2};n\widehat{p},n(1-\widehat{p})+1\right), \text{Upper: Beta}\left(1-{\alpha}/{2};n\widehat{p}+1,n(1-\widehat{p})\right),\\[0.1cm]
\nonumber &\text{where Beta} (\cdot) \text{ is the quantile function of the Beta distribution}. \\[0.1cm]
\nonumber \text{WS:\quad} & \dfrac{\widehat{p}+{z_{\alpha/2}^2}/{2n}\pm z_{\alpha/2}\sqrt{{\widehat{p}(1-\widehat{p})}/{n}+{z_{\alpha/2}^2}/{4n^2}}}{1+{z_{\alpha/2}^2}/{n}}.\\[0.1cm]
\nonumber \text{AC:\quad} & \widetilde{p}\pm z_{\alpha/2}\sqrt{\dfrac{\widetilde{p}(1-\widetilde{p})}{\widetilde{n}}}, \text{where } \widetilde{p}=\dfrac{n\widehat{p}+{z_{\alpha/2}^2}/{2}}{n+z_{\alpha/2}^2} \text{ and } \widetilde{n}=n+z_{\alpha/2}^2.
\end{align}


\subsection{\textit{Estimating Proportions when No Events are Observed}}
\label{sec:zeroevent}
\noindent To provide an initial flavour of the challenging nature of estimating rare-event probabilities, we first consider the situation where no events are observed in the sample. Letting $x$ denote the number of observed events, for very small $p$, it will be quite likely in small samples that $x=0$, and, hence, $\widehat{p}=0/n=0$. For example, consider an event that occurs with probability $p=10^{-3}$, and consider the case where the sample size is $n \le 100$ such that $\Pr(X = 0) > 0.9$, i.e., it is very likely that no events will be observed in this sample.

In this scenario where no events are observed, the resulting intervals are very conservative for small $n$. For example, with a sample of size $n=100$, the following intervals are obtained: W: $[0,0]$, CP: $[0, 0.0362]$, WS: $[0,0.0370]$, AC:$[-0.0074, 0.0444]$. Clearly the Wald interval is degenerate, and the remaining intervals are far too wide to be useful in small $p$ settings where a good estimate of magnitude is required; the issue is more acute for smaller proportions, e.g., the probability of observing no events is 0.99 if $p = 10^{-4}$ and 0.999 if $p = 10^{-5}$.  As it is clear that a sample size of $n=100$ will not suffice, further guidance is required in these scenarios, which we provide in the sequel. Of course, the required precision is analysis specific: the above intervals would be perfectly adequate if one was only interested in assessing if $p<0.1$, for example. However, for situations where it is important to determine the order of magnitude, such as assessing the failure rate in high-volume manufacturing, then sufficiently large samples will be required to gain a reasonable estimate of the order of magnitude of $p$.


\section{Evaluation Criteria}
\label{sec:Criteria}

The most commonly used CI evaluation criteria are coverage probability and expected width \citep{Goncalves:2012} --- and these are the criteria that we consider in this work. However, other performance metrics have been proposed. For example, \cite{Vos:2005} interpret an interval as the non-rejected parameter values in a hypothesis test and discuss the \textit{$p$-confidence} and \textit{$p$-bias} criteria; \cite{Newcombe:1998} presents a criterion using noncoverage as an indicator of location; and \cite{Park:2019} adopt an ensemble approach and use root mean squared error and mean absolute deviation to measure CI performance. 


\subsection{\textit{Coverage Probability}}

The coverage probability can be interpreted as the computed interval's long-run percentage inclusion of the unknown parameter. Denoting $L_x$ and $U_x$ as the lower and upper CI bounds formed with $x$ successes (suppressing the dependence on $p$ and the significance level, $\alpha$), the expected coverage probability, which we denote $CPr$, for a fixed parameter $p$, is given by

\bne \label{eq:CPr} CPr(n,p)=\sum_{x=0}^{n}{n \choose x}p^x(1-p)^{n-x}1(L_x\le p \le U_x),\ene

\noindent where $1(\cdot)$ is an indicator function that takes the value 1 when its argument is true, and 0 otherwise.


\subsection{\textit{Expected Width}}

The expected width, which we denote $EW$, is given by

\bne \label{eq:EW} EW(n,p)=\sum_{x=0}^{n}{n \choose x}p^x(1-p)^{n-x}(U_x-L_x),\ene

\noindent and the expected margin of error, $EMoE$, is then given as $EMoE(n,p)=EW(n,p)/{2}$.


\section{Calculating Sample Size}
\label{sec:Sample}

The first problem in CI estimation is determining the sample size required to achieve a desired estimation precision. There have been a range of sample size determination methods discussed in the literature, e.g., \cite{Goncalves:2012}, \cite{Korn:1986} and \cite{Liu:2002}, but here we adopt the common approach of deriving the sample size from the CI formula with fixed $\epsilon_R=\epsilon/p$. Used in conjunction with the Wald margin of error, one obtains

\bne \label{eq:Waldn}n=\left\lceil\dfrac{z_{\alpha/2}^2(1-p^*)}{{\epsilon_R}^2p^*}\right\rceil,\ene

\noindent where $\lceil \cdot \rceil$ denotes the ceiling function, and $p^*$ denotes an anticipated value of $p$, i.e., an initial estimate. (Sample size formulas for the Clopper-Pearson, Wilson and Agresti-Coull intervals are given in Appendix \ref{app:A}.)


\subsection{\textit{Initial Estimate of} \texorpdfstring{$\bm{p}$}{$p$}}
\label{sec:InitialEst}

Selecting a value for $p^*$ is required to make equation (\ref{eq:Waldn}) operational, and this is an inherent practical challenge in any such sample size calculation. In some situations, it might be possible to overcome this problem by utilizing subject matter knowledge or results from a previous study. If no previous information is available, a common approach is to consider the value of $p^*=0.5$, but we do not adopt this approach here given that the focus of this work is on small/rare-event probabilities.

In some situations one might be able to gain a reasonable estimate of $p$. Consider a manufacturing environment where, for example, past experience or consultation with process experts could provide an analyst with information on the order of magnitude of $p$. One may be able to deduce that the true proportion is more likely to be of the order 1 in 10,000 rather than 1 in 1,000. Such insight could be sufficient in setting a reasonable value for $p^*$ and subsequently determining an appropriate sample size. Such initial estimation of an unknown parameter is a topic worthy of further discussion but is beyond the scope of this work. Here the focus is on the performance of the intervals after an initial estimate has been obtained.


\subsection{\textit{Margin of Error Relative to} \texorpdfstring{$\bm{p}$}{$p$}}
\label{sec:MoE}

The required precision is largely analysis dependent; what could be considered reasonable accuracy in one setting might be completely inappropriate in another. Although a relative margin of error scheme cannot be rigidly prescribed, we suggest a general scheme to avoid intervals that are too wide to be practically useful, or indeed intervals that are too narrow, in the sense that a reasonable estimate could have been obtained using fewer resources.

To ensure that the margin of error is not larger than the order of magnitude of $p^*$, we impose $\epsilon_R\le1$. Thus, one could consider $0\le\epsilon_R\le1$ as a plausible margin of error scheme. However, considering $\epsilon_R$ values too close to the bound of 1 results in very wide intervals,  whilst considering $\epsilon_R$ values too close to the bound of 0 results in very narrow intervals. As $\epsilon_R$ approaches zero, an increasingly large sample size is required, and such high precision is not likely to be required for many studies. Therefore, we suggest $\epsilon_R \in [0.1, 0.5]$ as a reasonable scheme. This scheme ensures that the interval is not impractically wide, nor excessively narrow (in terms of demanding very large sample sizes), and we show in Section \ref{sec:CITable} that acceptable coverage is achieved for this range of $\epsilon_R$ values.\\

\noindent  A comparison of calculated sample sizes corresponding to $\epsilon_R=0.4$ is provided in Table \ref{tab:ncompare}. (Sample size values in this and subsequent tables are rounded to $2$ significant digits.) We can see from Table \ref{tab:ncompare} that the Wald, Wilson and Agresti-Coull sample sizes are similar across the $p^*$ range, but the Clopper-Pearson sample sizes are approximately $40\%$ larger which is indicative of this method's conservatism. Note that Table \ref{tab:ncompare} is only intended to provide an initial sense of the sample size requirements. However, final results can be found in Tables \ref{tab:CIperf1} and \ref{tab:CIperf2}, and these account for the fact that empirical coverage for binomial proportion interval estimators is non-monotonic in the sample size (see Sections \ref{sec:RelMoE} and \ref{sec:CITable}).

\begin{table}[!h]
\footnotesize
\centering
\caption{Sample size comparison for $\epsilon_R=0.4$}
\label{tab:ncompare}
\begin{tabular}{cccccc} 
\\[-0.6cm]
\hline\\[-0.35cm]
 ${{p^*}}$&${\epsilon}$&${\text{W}}$&${\text{CP}}$&${\text{WS}}$&${\text{AC}}$\\
\hline\\[-0.35cm]
$10^{-1}$&$4.0\cdot10^{-2}$&$2.2\cdot10^{2}$&$3.0\cdot10^{2}$&$2.2\cdot10^{2}$&$2.3\cdot10^{2}$\\
&&&&\\[-0.425cm]
$10^{-2}$&$4.0\cdot10^{-3}$&$2.4\cdot10^{3}$&$3.3\cdot10^{3}$&$2.5\cdot10^{3}$&$2.5\cdot10^{3}$\\
&&&&\\[-0.425cm]
$10^{-3}$&$4.0\cdot10^{-4}$&$2.4\cdot10^{4}$&$3.4\cdot10^{4}$&$2.5\cdot10^{4}$&$2.6\cdot10^{4}$\\
&&&&\\[-0.425cm]
$10^{-4}$&$4.0\cdot10^{-5}$&$2.4\cdot10^{5}$&$3.4\cdot10^{5}$&$2.5\cdot10^{5}$&$2.6\cdot10^{5}$\\
&&&&\\[-0.425cm]
$10^{-5}$&$4.0\cdot10^{-6}$&$2.4\cdot10^{6}$&$3.4\cdot10^{6}$&$2.5\cdot10^{6}$&$2.6\cdot10^{6}$\\
&&&&\\[-0.425cm]
$10^{-6}$&$4.0\cdot10^{-7}$&$2.4\cdot10^{7}$&$3.4\cdot10^{7}$&$2.5\cdot10^{7}$&$2.6\cdot10^{7}$\\
\hline\\[-0.45cm]
\end{tabular}
\end{table}

\noindent Given that $\epsilon_R$ is a function of $p^*$, the above calculated sample sizes are only applicable to each specific $p^*$. Even if the true proportion $p$ equals $p^*$ (i.e., the initial estimate is perfect), $\widehat{p}$ will of course vary from sample to sample and may not equal $p$; hence, the realized relative margin of error will typically differ from $\epsilon_R$. 


\subsection{\texorpdfstring{$\bm{\epsilon \text{-}p^*}$}{$\epsilon,p$} \textit{Compatibility}}
\label{sec:eps_p_compatibility}

Next we illustrate the importance of defining the margin of error in relation to the magnitude of the proportion. Consider the following fixed margin of error schemes:
\begin{multicols}{2}
\begin{itemize}
\item \hspace{0.25cm}Scheme $1$: $\epsilon=4\cdot10^{-2},\hspace{0.1cm} {p^*}={p}$\\[-0.75cm]
\item \hspace{0.25cm}Scheme $2$: $\epsilon=4\cdot10^{-2},\hspace{0.1cm} {p^*}=0.5$
\end{itemize}
\columnbreak
\begin{itemize}
\item \hspace{0.25cm}Scheme $3$: $\epsilon=4\cdot10^{-4},\hspace{0.1cm} {p^*}={p}$\\[-0.75cm]
\item \hspace{0.25cm}Scheme $4$: $\epsilon=4\cdot10^{-4},\hspace{0.1cm} {p^*}={0.5}$
\end{itemize}
\end{multicols}

\noindent Table \ref{tab:fixedmoe} displays the calculated Wald sample sizes and coverage probabilities corresponding to the above margin of error schemes. (A comparison of Wald, Clopper-Pearson, Wilson and Agresti-Coull coverage probabilities for the above $\epsilon$ schemes is given in Appendix \ref{app:B}.) Referring to Table \ref{tab:fixedmoe}, fixing $\epsilon=4\cdot10^{-2}$ and considering $p^*=p$ (\textit{Scheme 1}) creates sample sizes that reduce dramatically as a function of $p$. This results in coverage probabilities that are completely inadequate for $p \le 10^{-2}$. In \textit{Scheme 2}, both $\epsilon$ and $p^*$ are fixed and this creates a constant sample size of $n=6\cdot10^2$. This sample size is reasonable for $p=10^{-1}$, but is insufficient for the remaining $p$ values, which is reflected in the poor coverage performance. 

\textit{Scheme 3} is similar to \textit{Scheme 1}, but here, $\epsilon$ is reduced to $4\cdot10^{-4}$. This $\epsilon \text{-} p$ combination produces sufficient coverage for $p \ge 10^{-2}$, but deteriorates for the smaller $p$ values. In \textit{Scheme 4}, $p^*$ is fixed at $0.5$ and $\epsilon=4\cdot10^{-4}$, this results in a constant sample size of $n=6\cdot10^6$, which produces good coverage throughout the $p$ range, particularly for $p\ge 10^{-5}$. Whilst the coverage is satisfactory in this scheme, the magnitude of $\epsilon$ is not compatible with all $p$ values, particularly $p=10^{-1}$ and $p\le10^{-5}$. For $p=10^{-1}$ the resulting interval is $[0.0996,0.1004]$ which is too narrow in the sense that a reasonable interval could be obtained with a significantly reduced sample size. For $p=10^{-6}$ the interval is truncated at $[0,0.000401]$. Here, even though the coverage is reasonable, the interval is too wide to be practically useful since its width is two orders of magnitude larger than $p$. 

\begin{table}[!h]
\footnotesize
\centering
\caption{Wald-based sample size comparison - fixed $\epsilon$}
\label{tab:fixedmoe}
\begin{tabular}{ccccc} 
\\[-0.6cm]
\hline\\[-0.35cm]
\multirow{4}{*}{${p}$}&\multicolumn{4}{c}{Margin of Error Scheme}\\
&1&2&3&4\\
&$\epsilon=4\cdot10^{-2}$&$\epsilon=4\cdot10^{-2}$&$\epsilon=4\cdot10^{-4}$&$\epsilon=4\cdot10^{-4}$\\
&$p^*=p$&$p^*=0.5$&$p^*=p$&$p^*=0.5$\\
\hline\\[-0.35cm]
\multirow{2}{*}{$10^{-1}$}&$2.2\cdot10^{2}$&$6.0\cdot10^{2}$&$2.2\cdot10^{6}$&$6.0\cdot10^{6}$\\&(93.8)&(94.5)&(95.0)&(95.0)\\
&&&&\\[-0.425cm]
\multirow{2}{*}{$10^{-2}$}&$2.4\cdot10^{1}$&$6.0\cdot10^{2}$&$2.4\cdot10^{5}$&$6.0\cdot10^{6}$\\&(21.4)&(93.1)&(95.0)&(95.0)\\
&&&&\\[-0.425cm]
\multirow{2}{*}{$10^{-3}$}&$3.0\cdot10^{0}$&$6.0\cdot10^{2}$&$2.4\cdot10^{4}$&$6.0\cdot10^{6}$\\&(0.3)&(45.2)&(93.1)&(95.0)\\
&&&&\\[-0.425cm]
\multirow{2}{*}{$10^{-4}$}&$1.0\cdot10^{0}$&$6.0\cdot10^{2}$&$2.4\cdot10^{3}$&$6.0\cdot10^{6}$\\&(0)&(5.8)&(21.3)&(94.9)\\
&&&&\\[-0.425cm]
\multirow{2}{*}{$10^{-5}$}&$1.0\cdot10^{0}$&$6.0\cdot10^{2}$&$2.4\cdot10^{2}$&$6.0\cdot10^{6}$\\&(0)&(0.6)&(0.2)&(94.9)\\
&&&&\\[-0.425cm]
\multirow{2}{*}{$10^{-6}$}&$1.0\cdot10^{0}$&$6.0\cdot10^{2}$&$2.5\cdot10^{1}$&$6.0\cdot10^{6}$\\&(0)&(0.1)&(0)&(93.4)\\
&&&&\\[-0.4cm]
\hline
\multicolumn{5}{l}{\scriptsize Wald $95\%$ CI coverage shown in parentheses}\\[-0.1cm]
\multicolumn{5}{l}{\scriptsize Coverage computed with $p=p^*$}
\end{tabular}
\end{table}

\noindent The Wald sample sizes and coverage probabilities associated with the relative margin of error schemes: $\epsilon_R \in \{0.05, 0.1,0.2,0.3,0.4,0.5,0.75\}$ are given in Table \ref{tab:varmoe}. 

\begin{table}[!h]
\footnotesize
\centering
\caption{Wald-based sample size comparison - variable $\epsilon$}
\label{tab:varmoe}
\begin{tabular}{cccccccc} 
\\[-0.6cm]
\hline\\[-0.5cm]
\multirow{2}{*}{${p^*}$}&\multicolumn{7}{c}{$\epsilon_R$}\\
&0.05&0.1&0.2&0.3&0.4&0.5&0.75\\[-0.05cm]
\hline\\[-0.4cm]
\multirow{2}{*}{$10^{-1}$}&$1.4\cdot10^{4}$&$3.5\cdot10^{3}$&$8.6\cdot10^{2}$&$3.8\cdot10^{2}$&$2.2\cdot10^{2}$&$1.4\cdot10^{2}$&$6.2\cdot10^{1}$\\&(94.9)&(94.9)&(95.0)&(94.3)&(93.8)&(93.3)&(94.7)\\
&&&&&&&\\[-0.425cm]
\multirow{2}{*}{$10^{-2}$}&$1.5\cdot10^{5}$&$3.8\cdot10^{4}$&$9.5\cdot10^{3}$&$4.2\cdot10^{3}$&$2.4\cdot10^{3}$&$1.5\cdot10^{3}$&$6.8\cdot10^{2}$\\&(94.9)&(94.9)&(95.2)&(94.0)&(95.2)&(92.5)&(90.2)\\
&&&&&&&\\[-0.425cm]
\multirow{2}{*}{$10^{-3}$}&$1.5\cdot10^{6}$&$3.8\cdot10^{5}$&$9.6\cdot10^{4}$&$4.3\cdot10^{4}$&$2.4\cdot10^{4}$&$1.5\cdot10^{4}$&$6.8\cdot10^{3}$\\&(94.9)&(95.0)&(95.0)&(94.7)&(93.1)&(93.3)&(90.4)\\
&&&&&&&\\[-0.425cm]
\multirow{2}{*}{$10^{-4}$}&$1.5\cdot10^{7}$&$3.8\cdot10^{6}$&$9.6\cdot10^{5}$&$4.3\cdot10^{5}$&$2.4\cdot10^{5}$&$1.5\cdot10^{5}$&$6.8\cdot10^{4}$\\&(95.0)&(95.0)&(95.0)&(94.7)&(93.1)&(93.3)&(90.4)\\
&&&&&&&\\[-0.425cm]
\multirow{2}{*}{$10^{-5}$}&$1.5\cdot10^{8}$&$3.8\cdot10^{7}$&$9.6\cdot10^{6}$&$4.3\cdot10^{6}$&$2.4\cdot10^{6}$&$1.5\cdot10^{6}$&$6.8\cdot10^{5}$\\&(95.0)&(95.0)&(95.0)&(94.7)&(93.1)&(93.3)&(90.4)\\
&&&&&&&\\[-0.425cm]
\multirow{2}{*}{$10^{-6}$}&$1.5\cdot10^{9}$&$3.8\cdot10^{8}$&$9.6\cdot10^{7}$&$4.3\cdot10^{7}$&$2.4\cdot10^{7}$&$1.5\cdot10^{7}$&$6.8\cdot10^{6}$\\&(95.0)&(95.0)&(95.0)&(94.7)&(93.1)&(93.3)&(90.4)\\
&&&&&&&\\[-0.4cm]
\hline
\multicolumn{8}{l}{\scriptsize Wald $95\%$ CI coverage shown in parentheses}\\[-0.1cm]
\end{tabular}
\end{table}

From Table \ref{tab:varmoe}, we see that by considering $\epsilon$ in relation to the magnitude of $p$, the coverage probabilities are reasonable across the $p$ range, but now, the analyst must choose a scheme such that the resulting interval's width is appropriate. For example, consider $p^*=p=10^{-1}$ and $\epsilon_R=0.05$ where the resulting interval is $[0.095,0.105]$. This interval is very narrow and the large sample size of $1.4\cdot10^{4}$ reflects this quite stringent margin of error. Moving to $\epsilon_R=0.75$ has the advantage of significantly reducing the sample size, but, of course, the interval is significantly wider at $[0.025,0.175]$. To obtain intervals that are neither too liberal nor too conservative, that are reasonable in terms of coverage performance, and which avoid excessively large sample sizes, we recommend $\epsilon_R \in [0.1,0.5]$ as a reasonable scheme.

The coverage values for the Clopper-Pearson, Wilson and Agresti-Coull intervals are similar to those shown in Table \ref{tab:varmoe} for $\epsilon_R\le0.5$. A comparison of coverage probabilities for $\epsilon_R=0.75$ is given in Appendix \ref{app:B}.


\subsection{\textit{Suitability of} \texorpdfstring{$\bm{\epsilon_R}$}{$\epsilon,R$} \textit{Scheme}}
\label{sec:MoEScheme}

A range of qualifications/criteria are often used to check the validity of using approximate CI estimators. \cite{Fleiss:2003} state that the normal distribution provides excellent approximations to exact binomial procedures when $np\ge5$ and $n(1-p)\ge5$. \cite{Leemis:1996} also discuss the $np\ge5 \hspace{0.1cm}(\text{or }10)$ and $n(1-p)\ge 5 \hspace{0.1cm}(\text{or }10)$ qualification.

We examine the proposed $\epsilon_R$ scheme to assess its compatibility with the qualification $np^*\ge a$ and $n(1-p^*)\ge a$, where $a\in\{5,10\}$, in relation to the Wald sample size equation.

\noindent Multiplying equation (\ref{eq:Waldn}) by $p^*$ gives

\bne \label{eq:np} np^*=\dfrac{z_{\alpha/2}^2(1-p^*)}{{\epsilon_R}^2}\ge a\implies \epsilon_R \le\sqrt{\dfrac{z_{\alpha/2}^2(1-p^*)}{a}}.\ene

\noindent For a given $n$, $np^*<n(1-p^*)$ when $p^*<0.5$ and, therefore, equation (\ref{eq:np}) is sufficient in the small-$p$ regime to ensure both $np^*$ and $n(1-p^*)$ are greater than $a$. An evaluation of $\epsilon_R$ for $p^*\in [10^{-6},10^{-1}]$ and $\alpha \in \{0.1,0.05,0.01\}$ is provided in Table \ref{tab:np5}, which shows that our suggested relative margin of error scheme, $\epsilon_R\in[0.1,0.5]$, lies below the threshold of equation (\ref{eq:np}). (The case of $\epsilon_R=0.5$ negligibly exceeds the threshold of $0.493$ for $p^*=10^{-1}, \alpha=0.1$.) 

\begin{table}[!h]
\footnotesize
\centering
\caption{$\epsilon_R$ thresholds as per equation (\ref{eq:np}) for $a \in \{5,10\}$}
\label{tab:np5}
\begin{tabular}{ccccccc} 
\\[-0.55cm]
\hline\\[-0.4cm]
\multirow{2}{*}{${p^*}$}&\multicolumn{3}{c}{$a=5$}&\multicolumn{3}{c}{$a=10$}\\
&$\alpha=0.1$&$\alpha=0.05$&$\alpha=0.01$&$\alpha=0.1$&$\alpha=0.05$&$\alpha=0.01$\\
\hline\\[-0.35cm]
$10^{-1}$&0.698&0.832&1.093&0.493&0.588&0.773\\
&&&\\[-0.4225cm]
$10^{-2}$&0.732&0.872&1.146&0.518&0.617&0.810\\
&&&\\[-0.45cm]
$10^{-3}$&0.735&0.876&1.151&0.520&0.619&0.814\\
&&&\\[-0.425cm]
$10^{-4}$&0.736&0.876&1.152&0.520&0.620&0.815\\
&&&\\[-0.425cm]
$10^{-5}$&0.736&0.877&1.152&0.520&0.620&0.815\\
&&&\\[-0.425cm]
$10^{-6}$&0.736&0.877&1.152&0.520&0.620&0.815\\
&&&\\[-0.425cm]
\hline
\end{tabular}
\end{table}


\subsection{\textit{Tolerances for Assessing CI Performance}}
\label{sec:Tolerances}

In this section we suggest suitable tolerances for assessing interval performance in terms of coverage probability and relative margin of error. In relation to achieving a desired coverage probability, one usually considers $(1-\alpha)100\pm\epsilon^*\%$, where $\epsilon^*$ denotes a predefined coverage tolerance. The definition of such a tolerance is dependent on the individual researcher and particular study, and is thus difficult to quantify. In one study $(1-\alpha)100\pm4\%$ might be acceptable, whilst in another, one might require $(1-\alpha)100\pm0.5\%$. We suggest that $\epsilon^*\in\{1,2,3\}$ would be reasonable tolerances for most analyses, and as such, consider acceptable expected coverage probabilities as $CPr\in (1-\alpha)100\pm3\%$, where $CPr$ is described in equation (\ref{eq:CPr}).

A tolerance is also necessary with regard to the relative margin of error. As with the coverage, the desired margin of error is dependent on the particular research question and hence can not be rigidly prescribed. However, as previously discussed, it is important that the magnitude of the margin of error reflect the magnitude of the estimated proportion.\\

\noindent Table \ref{tab:tol} provides suggested tolerances for the assessment of $\epsilon_R$ and $CPr$ for $(1-\alpha)100\%$ confidence intervals which could be considered reasonable in most settings.

\begin{table}[!h]
\footnotesize
\centering
\caption{Coverage \& relative margin of error tolerances}
\label{tab:tol}
\begin{tabular}{lcr} 
\\[-0.55cm]
\hline\\[-0.35cm]
{Tolerance}&{Coverage}&{Relative MoE}\\
\hline\\[-0.35cm]
Target&\hspace{0.75cm}$CPr\in(1-\alpha)100\pm1\%$&$\epsilon_R\le0.5$\\
&&\\[-0.425cm]
Acceptable&\hspace{0.75cm}$CPr\in(1-\alpha)100\pm2\%$&$\epsilon_R\le0.75^*$\\
&&\\[-0.425cm]
Minimally acceptable&\hspace{0.75cm}$CPr\in(1-\alpha)100\pm3\%$&$\epsilon_R\le1$\\
&&\\[-0.425cm]
Unacceptable&\hspace{0.75cm}$CPr\notin(1-\alpha)100\pm3\%$&$\epsilon_R>1$\\
\hline
\multicolumn{3}{l}{\scriptsize $^*$Considered \textit{acceptable} as a compromise between $\epsilon_R\le0.5$ (desired) and $\epsilon_R\le1$ (limit)}
\end{tabular}
\end{table}


\section{Relative Margin of Error Central to Performance}
\label{sec:RelMoE}

Next we illustrate how the relative margin of error is fundamental to CI performance evaluation. We show that when a valid confidence interval is defined as achieving a desired coverage probability whilst simultaneously satisfying a minimum relative margin of error, the four interval estimators perform similarly for a given $n\text{-}p^*\text{-}\alpha$ combination. 

To demonstrate CI performance we consider the \textit{expected} relative margin of error, which we define as $\tilde{\epsilon}_R = {EMoE}/{p},$ where $EMoE$ is half of the expected width (see equation (\ref{eq:EW})). As discussed in Section \ref{sec:MoE}, a relative margin of error exceeding $1$ is not acceptable from a practical perspective, and, as per Section \ref{sec:MoEScheme}, we suggest that it should not exceed $0.5$.

\begin{figure}[!h]
\centering
\includegraphics[width=16cm, height=11cm]{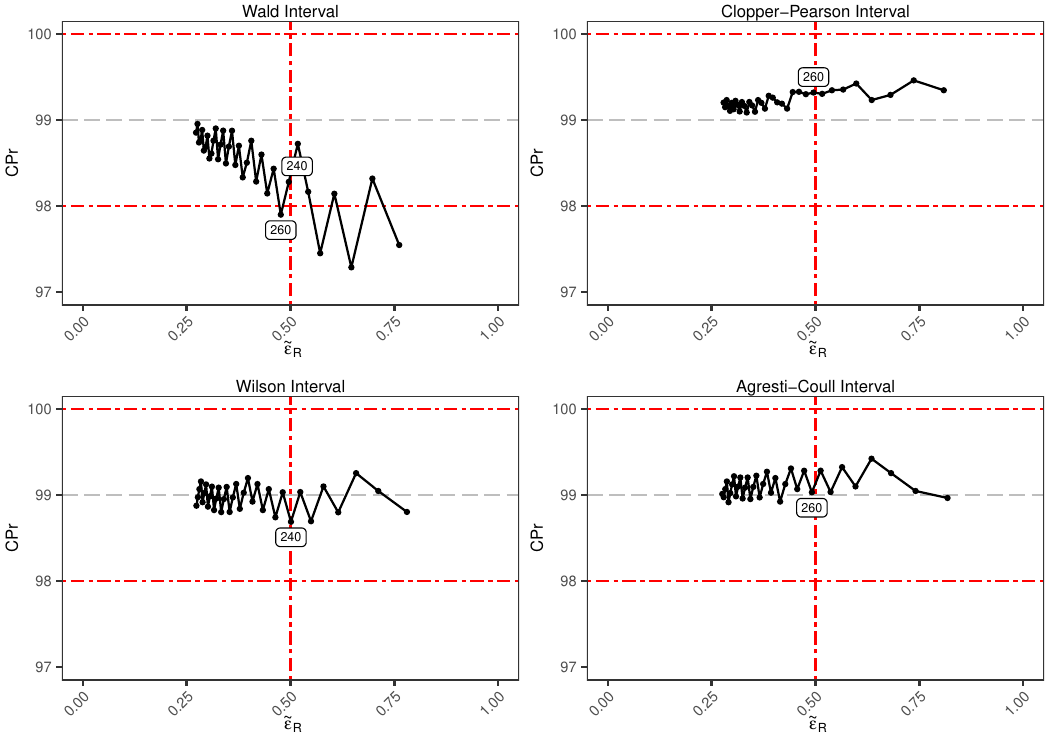}
\caption{$CPr$ versus $\tilde{\epsilon}_R$ for ${p^*}=p=10^{-1}, \alpha=0.01$. Dashed (grey) line represents the nominal $CPr$ value. Dot-dashed (red) lines represent the target $CPr$ and $\tilde{\epsilon}_R$ tolerances from Table \ref{tab:tol}. Sample size range is from $n=100$, to $n=800$, in steps of $20$. Labels shown adjacent to data points depict the first $n$ where both $CPr$ and $\tilde{\epsilon}_R$ requirements are satisfied. Additional data labels are referred to within the text.}
\label{Fig1}
\end{figure}

\begin{figure}[!h]
\centering
\includegraphics[width=16cm, height=11cm]{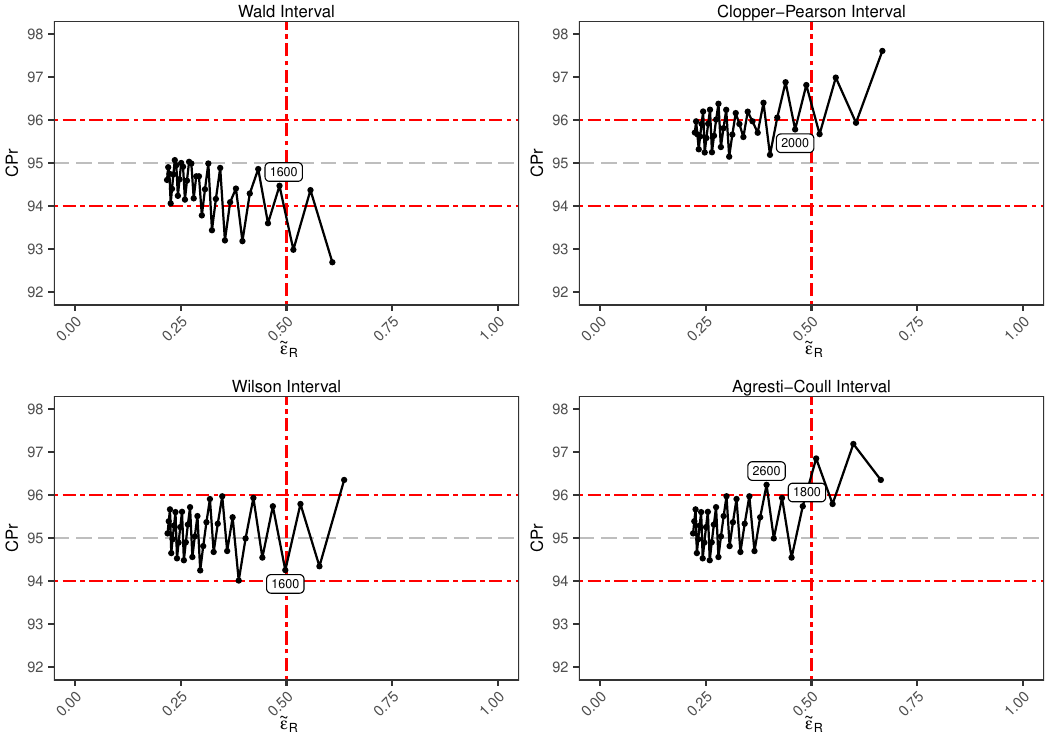}
\caption{$CPr$ versus $\tilde{\epsilon}_R$ for ${p^*}=p=10^{-2},\alpha=0.05$. Dashed (grey) line represents the nominal $CPr$ value. Dot-dashed (red) lines represent the target $CPr$ and $\tilde{\epsilon}_R$ tolerances from Table \ref{tab:tol}. Sample size range is from $n=$ 1,000, to $n=$ 8,000, in steps of $200$. Labels shown adjacent to data points depict the first $n$ where both $CPr$ and $\tilde{\epsilon}_R$ requirements are satisfied. Additional data labels are referred to within the text.}
\label{Fig2}
\end{figure}

\noindent Figure \ref{Fig1} provides a $99\%$ CI performance comparison for $p^*=p=10^{-1}$, and shows that the Wilson, Agresti-Coull and Clopper-Pearson intervals all achieve satisfactory coverage across the sample size range, whereas for $n\le260$, the coverage of the Wald interval oscillates around the lower limit of $98\%$. For example, the coverage is satisfactory at $n=240$, but then falls below $98\%$ at $n=260$. This phenomenon of coverage oscillation relates to the discreteness of the binomial distribution and has been previously discussed in the literature, e.g., \cite{Agresti:1998}, \cite{Andersson:2023}, \cite{Blyth:1983}, \cite{Brown:2001}, and \cite{Vollset:1993}. For a given $p$, the empirical coverage does converge to the $(1-\alpha)100\%$ level with $n$ as one would expect, but it does so in an oscillatory fashion for neighbouring values of $n$. Figures \ref{Fig1} and \ref{Fig2} show that all four estimators suffer from this erratic behaviour.

Whilst the coverage performance of the Wald interval is inferior to the other three intervals for $n<240$, none of the intervals satisfy the $\tilde{\epsilon}_R\le0.5$ requirement at these lower sample sizes. Thus, by stipulating a minimum requirement for $\tilde{\epsilon}_R$, the poor coverage at small $n$ is rendered irrelevant and the performance of the Wald interval is more comparable to the other three intervals when $\tilde{\epsilon}_R\le0.5$.\\

\noindent A comparison of the performance of a $95\%$ CI for $p^*=p=10^{-2}$ is given in Figure \ref{Fig2} which shows that the Wald interval achieves $\tilde{\epsilon}_R\le0.5$ for $n\ge$ 1,600. For $n\ge$ 1,600 the Wald interval encounters five sample sizes where the coverage drops below the lower limit of $94\%$. The Clopper-Pearson interval requires a sample size of $n=$ 2,000 to satisfy both $\tilde{\epsilon}_R\le0.5$ and $CPr \in [94\%,96\%]$, with the coverage exceeding the upper limit of $96\%$ on seven occasions for $n >$ 2,000. The Agresti-Coull interval performs very well for $n\ge1,800$, with just one value ($n=$ 2,600), failing to satisfy both $CPr$ and $\tilde{\epsilon}_R$ thereafter. The Wilson interval provides the best performance, satisfying both $CPr$ and $\tilde{\epsilon}_R$ requirements $\forall \hspace{0.1cm}n\ge$ 1,600.

Figures \ref{Fig1} and \ref{Fig2} highlight the similarities in performance when one considers $\tilde{\epsilon}_R$. In general, moderate-to-large sample sizes are required to satisfy both $CPr$ and $\tilde{\epsilon}_R$ criteria, and at these sample sizes the performance across the four intervals is reasonably comparable.


\section{CI Performance Tables}
\label{sec:CITable}

Tables \ref{tab:p10-1_smalln} through \ref{tab:p10-6_largen} provide a $95\%$ CI comparison for $p^*=p=10^{-1}$ and $p^*=p=10^{-6}$, across a range of sample sizes and further illustrate the performance similarities among the estimators. The table cells are colour coded according to the tolerances discussed in Table \ref{tab:tol}: target (green), acceptable (yellow), minimally acceptable (orange) and unacceptable (red).

We first consider $p^*=p=10^{-1}$ and $10\le n\le140$, with Table \ref{tab:p10-1_smalln} showing that none of the intervals satisfy the desired $CPr$ and $\tilde{\epsilon}_R$ requirements simultaneously. The importance of considering the relative margin of error in CI evaluation is clearly evident. In several cases the coverage probability lies within the desired tolerance but the excessive relative margin of error renders the estimate impractical. For example, referring to the Wilson interval, $CPr(20,10^{-1})=95.7\%$, however, $\tilde{\epsilon}_R=1.32$ which is not acceptable.

\begin{table}[!h]
\footnotesize
\centering
\caption{$95\%$ CI performance - ${p^*}=p=10^{-1}$, ``small'' $n$}
\label{tab:p10-1_smalln}
\begin{tabular}{ccccccccc} 
\\[-0.4cm]
\hline\\[-0.4cm]
\multirow{2}{*}{${n}$}&\multicolumn{2}{c}{{W}}&\multicolumn{2}{c}{{CP}}&\multicolumn{2}{c}{{WS}}&\multicolumn{2}{c}{{AC}}\\
&{${CPr}$}&${\tilde{\epsilon}_R}$&{${CPr}$}&${\tilde{\epsilon}_R}$&{${CPr}$}&${\tilde{\epsilon}_R}$&{${CPr}$}&${\tilde{\epsilon}_R}$\\
\hline\\[-0.41cm]
10&\cre65.0&\cre1.40&\cre98.7&\cre2.12&\cy93.0&\cre1.85&\cy93.0&\cre2.11\\
&&&&&&&&\\[-0.475cm]
20&\cre87.6&\cre1.17&\cre98.9&\cre1.46&\cg95.7&\cre1.32&\cg95.7&\cre1.46\\
&&&&&&&&\\[-0.475cm]
30&\cre80.9&\cre1.01&\cre99.2&\cre1.18&\co97.4&\cre1.07&\co97.4&\cre1.16\\
&&&&&&&&\\[-0.475cm]
40&\cre91.4&\co0.89&\cy97.0&\cre1.01&\cg94.3&\co0.93&\cg95.8&\co0.99\\
&&&&&&&&\\[-0.475cm]
50&\cre87.9&\co0.81&\cy97.0&\co0.90&\cy97.0&\co0.83&\cy97.0&\co0.88\\
&&&&&&&&\\[-0.475cm]
60&\cg94.1&\cy0.74&\co97.2&\co0.82&\cg95.2&\co0.76&\cg95.2&\co0.80\\
&&&&&&&&\\[-0.475cm]
70&\co92.0&\cy0.69&\cg95.5&\co0.76&\cy93.2&\cy0.70&\co97.4&\cy0.73\\
&&&&&&&&\\[-0.475cm]
80&\cre90.0&\cy0.65&\co97.7&\cy0.71&\cy96.3&\cy0.66&\cy96.3&\cy0.68\\
&&&&&&&&\\[-0.475cm]
90&\cg94.6&\cy0.61&\cy96.7&\cy0.66&\cg95.0&\cy0.62&\cg95.0&\cy0.64\\
&&&&&&&&\\[-0.475cm]
100&\cy93.2&\cy0.58&\cg95.6&\cy0.63&\cy93.6&\cy0.59&\co97.2&\cy0.61\\
&&&&&&&&\\[-0.475cm]
110&\cre91.9&\cy0.55&\co97.6&\cy0.60&\cy96.3&\cy0.56&\cy96.3&\cy0.58\\
&&&&&&&&\\[-0.475cm]
120&\cg95.4&\cy0.53&\cy96.8&\cy0.57&\cg95.4&\cy0.54&\cg95.4&\cy0.55\\
&&&&&&&&\\[-0.475cm]
130&\cg94.4&\cy0.51&\cg96.0&\cy0.55&\cg94.5&\cy0.52&\cg94.5&\cy0.53\\
&&&&&&&&\\[-0.475cm]
140&\cy93.5&\cg0.49&\cy96.7&\cy0.53&\cy93.5&\cg0.50&\cy96.7&\cy0.51\\
&&&&&&&&\\[-0.425cm]
\hline
\end{tabular}
\end{table}

\noindent We see from Table \ref{tab:p10-1_largen} that by increasing $n$, $\tilde{\epsilon}_R \le 0.5$ is satisfied (at $n=150$, the Clopper-Pearson interval marginally exceeds $0.5$, but is less than $0.5$ thereafter). Each interval encounters sample sizes where $CPr$ exceeds the bounds of $\pm1\%$, but overall, $CPr$ is satisfactory. 

\begin{table}[!h]
\footnotesize
\centering
\caption{$95\%$ CI performance - ${p^*}=p=10^{-1}$, ``large'' $n$}
\label{tab:p10-1_largen}
\begin{tabular}{ccccccccc} 
\\[-0.41cm]
\hline\\[-0.4cm]
\multirow{2}{*}{${n}$}&\multicolumn{2}{c}{{W}}&\multicolumn{2}{c}{{CP}}&\multicolumn{2}{c}{{WS}}&\multicolumn{2}{c}{{AC}}\\
&{${CPr}$}&$\tilde{\epsilon}_R$&{${CPr}$}&$\tilde{\epsilon}_R$&{${CPr}$}&$\tilde{\epsilon}_R$&{${CPr}$}&$\tilde{\epsilon}_R$\\
\hline\\[-0.41cm]
150&\co92.6&\cg0.48&\co97.2&\cy0.51&\cg96.0&\cg0.48&\cg96.0&\cg0.49\\
&&&&&&&&\\[-0.475cm]
160&\cg95.5&\cg0.46&\cy96.6&\cg0.49&\cg95.4&\cg0.47&\cg95.4&\cg0.47\\
&&&&&&&&\\[-0.475cm]
170&\cg94.8&\cg0.45&\cg96.0&\cg0.48&\cg94.6&\cg0.45&\cg94.6&\cg0.46\\
&&&&&&&&\\[-0.475cm]
180&\cg94.1&\cg0.43&\cg95.4&\cg0.46&\cy93.9&\cg0.44&\cy96.6&\cg0.45\\
&&&&&&&&\\[-0.475cm]
190&\cy93.4&\cg0.42&\cy96.1&\cg0.45&\cy96.1&\cg0.43&\cy96.1&\cg0.43\\
&&&&&&&&\\[-0.475cm]
200&\co92.7&\cg0.41&\cy96.7&\cg0.44&\cg95.6&\cg0.42&\cg95.6&\cg0.42\\
&&&&&&&&\\[-0.475cm]
210&\cg95.3&\cg0.40&\cy96.2&\cg0.43&\cg95.1&\cg0.41&\cg95.1&\cg0.41\\
&&&&&&&&\\[-0.475cm]
220&\cg94.8&\cg0.39&\cg95.7&\cg0.42&\cg94.5&\cg0.40&\cg94.5&\cg0.40\\
&&&&&&&&\\[-0.475cm]
230&\cg94.3&\cg0.39&\cy96.4&\cg0.41&\cg94.0&\cg0.39&\cy96.4&\cg0.39\\
&&&&&&&&\\[-0.475cm]
240&\cy93.7&\cg0.38&\cg96.0&\cg0.40&\cg96.0&\cg0.38&\cg96.0&\cg0.38\\
&&&&&&&&\\[-0.475cm]
250&\cy93.2&\cg0.37&\cy96.6&\cg0.39&\cg95.6&\cg0.37&\cg95.6&\cg0.38\\
&&&&&&&&\\[-0.475cm]
260&\cg95.5&\cg0.36&\cy96.2&\cg0.38&\cg95.1&\cg0.36&\cg95.1&\cg0.37\\
&&&&&&&&\\[-0.475cm]
270&\cg95.0&\cg0.36&\cg95.8&\cg0.37&\cg94.7&\cg0.36&\cg94.7&\cg0.36\\
&&&&&&&&\\[-0.475cm]
280&\cg94.6&\cg0.35&\cg95.4&\cg0.37&\cg94.3&\cg0.35&\cg94.3&\cg0.36\\
&&&&&&&&\\[-0.475cm]
290&\cg94.2&\cg0.34&\cy96.1&\cg0.36&\cy96.1&\cg0.35&\cy96.1&\cg0.35\\
&&&&&&&&\\[-0.475cm]
300&\cy93.8&\cg0.34&\cy96.6&\cg0.35&\cg95.7&\cg0.34&\cg95.7&\cg0.34\\
&&&&&&&&\\[-0.475cm]
310&\cy93.3&\cg0.33&\cy96.3&\cg0.35&\cg95.4&\cg0.33&\cg95.4&\cg0.34\\
&&&&&&&&\\[-0.475cm]
320&\cg95.4&\cg0.33&\cg96.0&\cg0.34&\cg95.0&\cg0.33&\cg95.0&\cg0.33\\
&&&&&&&&\\[-0.475cm]
330&\cg95.1&\cg0.32&\cg95.7&\cg0.34&\cg94.7&\cg0.32&\cg94.7&\cg0.33\\
&&&&&&&&\\[-0.475cm]
340&\cg94.7&\cg0.32&\cg95.4&\cg0.33&\cg94.3&\cg0.32&\cg94.3&\cg0.32\\
&&&&&&&&\\[-0.475cm]
350&\cg94.4&\cg0.31&\cg96.0&\cg0.33&\cg96.0&\cg0.31&\cg96.0&\cg0.32\\
&&&&&&&&\\[-0.425cm]
\hline
\end{tabular}
\end{table}

\noindent As per Table \ref{tab:p10-6_smalln}, none of the intervals satisfy $CPr\in[94\%,96\%]$ and $\tilde{\epsilon}_R\le0.5$ for $p^*=p=10^{-6}$ and $n\le14\cdot10^{6}$. The Wilson method performs best in this scheme, and if the tolerances of $CPr\in[93\%,97\%]$ and $\tilde{\epsilon}_R\le0.75$ were considered, it would produce a valid interval $\forall \hspace{0.1cm} n$. \\

\begin{table}[!h]
\footnotesize
\centering
\caption{$95\%$ CI performance - ${p^*}=p=10^{-6}$, ``small'' $n$}
\label{tab:p10-6_smalln}
\begin{tabular}{ccccccccc} 
\\[-0.4cm]
\hline\\[-0.4cm]
\multirow{2}{*}{${n}$}&\multicolumn{2}{c}{{W}}&\multicolumn{2}{c}{{CP}}&\multicolumn{2}{c}{{WS}}&\multicolumn{2}{c}{{AC}}\\
&{${CPr}$}&$\tilde{\epsilon}_R$&{${CPr}$}&$\tilde{\epsilon}_R$&{${CPr}$}&$\tilde{\epsilon}_R$&{${CPr}$}&$\tilde{\epsilon}_R$\\
\hline\\[-0.41cm]
$7.5\cdot10^6$&\cy93.6&\cy0.70&\cg95.8&\co0.79&\cy93.7&\cy0.75&\co97.4&\co0.79\\
&&&&&&&&\\[-0.475cm]
$8.0\cdot10^6$&\cre89.2&\cy0.68&\cy96.9&\co0.76&\cg95.2&\cy0.72&\cg95.2&\co0.76\\
&&&&&&&&\\[-0.475cm]
$8.5\cdot10^6$&\cre91.9&\cy0.66&\co97.7&\cy0.73&\cy96.3&\cy0.70&\cy96.3&\cy0.74\\
&&&&&&&&\\[-0.475cm]
$9.0\cdot10^6$&\cg94.0&\cy0.64&\cg95.7&\cy0.71&\cy93.7&\cy0.68&\co97.2&\cy0.71\\
&&&&&&&&\\[-0.475cm]
$9.5\cdot10^6$&\cre90.3&\cy0.63&\cy96.7&\cy0.69&\cg95.2&\cy0.66&\cg95.2&\cy0.69\\
&&&&&&&&\\[-0.475cm]
$10.0\cdot10^6$&\co92.6&\cy0.61&\co97.5&\cy0.67&\cy96.3&\cy0.64&\cy96.3&\cy0.67\\
&&&&&&&&\\[-0.475cm]
$10.5\cdot10^6$&\cg94.4&\cy0.60&\cg95.7&\cy0.65&\cy93.9&\cy0.63&\co97.1&\cy0.65\\
&&&&&&&&\\[-0.475cm]
$11.0\cdot10^6$&\cre91.2&\cy0.58&\cy96.7&\cy0.64&\cg95.3&\cy0.61&\cg95.3&\cy0.64\\
&&&&&&&&\\[-0.475cm]
$11.5\cdot10^6$&\cy93.2&\cy0.57&\co97.5&\cy0.62&\cy96.3&\cy0.60&\cy96.3&\cy0.62\\
&&&&&&&&\\[-0.475cm]
$12.0\cdot10^6$&\cg94.3&\cy0.56&\cg95.8&\cy0.61&\cg94.2&\cy0.58&\co97.1&\cy0.60\\
&&&&&&&&\\[-0.475cm]
$12.5\cdot10^6$&\co92.1&\cy0.55&\cy96.8&\cy0.60&\cg95.5&\cy0.57&\cg95.5&\cy0.59\\
&&&&&&&&\\[-0.475cm]
$13.0\cdot10^6$&\cy93.8&\cy0.54&\co97.5&\cy0.58&\cy96.4&\cy0.56&\cy96.4&\cy0.58\\
&&&&&&&&\\[-0.475cm]
$13.5\cdot10^6$&\cg94.7&\cy0.53&\cg96.0&\cy0.57&\cg94.6&\cy0.55&\cg94.6&\cy0.57\\
&&&&&&&&\\[-0.475cm]
$14.0\cdot10^6$&\co92.9&\cy0.52&\cy96.9&\cy0.56&\cg95.7&\cy0.54&\cg95.7&\cy0.55\\
&&&&&&&&\\[-0.425cm]
\hline
\end{tabular}
\end{table}

\noindent Referring to Table \ref{tab:p10-6_largen}, for $15\cdot10^6 \le n \le 25\cdot10^6$ and $p^*=p=10^{-6}$, the Wald interval has the worst coverage, with four $CPr$ values exceeding $95 \pm2\%$. The Wilson and Agresti-Coull intervals perform the best, but overall, all four intervals perform well in this large sample size scheme, particularly if the coverage tolerance was considered as $CPr\in95\pm2\%$.\\

\begin{table}[!h]
\footnotesize
\centering
\caption{$95\%$ CI performance - ${p^*}=p=10^{-6}$, ``large'' $n$}
\label{tab:p10-6_largen}
\begin{tabular}{ccccccccc} 
\\[-0.4cm]
\hline\\[-0.4cm]
\multirow{2}{*}{${n}$}&\multicolumn{2}{c}{{W}}&\multicolumn{2}{c}{{CP}}&\multicolumn{2}{c}{{WS}}&\multicolumn{2}{c}{{AC}}\\
&{${CPr}$}&$\tilde{\epsilon}_R$&{${CPr}$}&$\tilde{\epsilon}_R$&{${CPr}$}&$\tilde{\epsilon}_R$&{${CPr}$}&$\tilde{\epsilon}_R$\\
\hline\\[-0.41cm]
$15.0\cdot10^6$&\cre91.9&\cg0.50&\cy96.3&\cy0.54&\cg94.9&\cy0.52&\cg94.9&\cy0.53\\
&&&&&&&&\\[-0.475cm]
$15.5\cdot10^6$&\cy93.6&\cg0.49&\co97.1&\cy0.53&\cg96.0&\cy0.51&\cg96.0&\cy0.52\\
&&&&&&&&\\[-0.475cm]
$16.0\cdot10^6$&\cg94.4&\cg0.49&\cg95.6&\cy0.52&\cg94.1&\cg0.50&\cy96.8&\cy0.52\\
&&&&&&&&\\[-0.475cm]
$16.5\cdot10^6$&\co92.7&\cg0.48&\cy96.5&\cy0.51&\cg95.3&\cg0.49&\cg95.3&\cy0.51\\
&&&&&&&&\\[-0.475cm]
$17.0\cdot10^6$&\cg94.2&\cg0.47&\co97.2&\cy0.51&\cy96.2&\cg0.49&\cy96.2&\cg0.50\\
&&&&&&&&\\[-0.475cm]
$17.5\cdot10^6$&\cg94.9&\cg0.47&\cg95.9&\cg0.50&\cg94.6&\cg0.48&\cg94.6&\cg0.49\\
&&&&&&&&\\[-0.475cm]
$18.0\cdot10^6$&\cy93.5&\cg0.46&\cy96.7&\cg0.49&\cg95.6&\cg0.47&\cg95.6&\cg0.48\\
&&&&&&&&\\[-0.475cm]
$18.5\cdot10^6$&\cg94.8&\cg0.45&\cg95.3&\cg0.48&\cy93.9&\cg0.46&\cy96.5&\cg0.48\\
&&&&&&&&\\[-0.475cm]
$19.0\cdot10^6$&\co92.8&\cg0.45&\cy96.2&\cg0.48&\cg95.0&\cg0.46&\cg95.0&\cg0.47\\
&&&&&&&&\\[-0.475cm]
$19.5\cdot10^6$&\cg94.1&\cg0.44&\cy97.0&\cg0.47&\cg96.0&\cg0.45&\cg96.0&\cg0.46\\
&&&&&&&&\\[-0.475cm]
$20.0\cdot10^6$&\cg94.8&\cg0.44&\cg95.7&\cg0.46&\cg94.4&\cg0.45&\cg94.4&\cg0.46\\
&&&&&&&&\\[-0.475cm]
$20.5\cdot10^6$&\cy93.5&\cg0.43&\cy96.5&\cg0.46&\cg95.5&\cg0.44&\cg95.5&\cg0.45\\
&&&&&&&&\\[-0.475cm]
$21.0\cdot10^6$&\cg94.7&\cg0.43&\cg95.1&\cg0.45&\cy93.8&\cg0.43&\cy96.3&\cg0.44\\
&&&&&&&&\\[-0.475cm]
$21.5\cdot10^6$&\co92.9&\cg0.42&\cg96.0&\cg0.45&\cg94.9&\cg0.43&\cg94.9&\cg0.44\\
&&&&&&&&\\[-0.475cm]
$22.0\cdot10^6$&\cg94.2&\cg0.42&\cy96.8&\cg0.44&\cg95.8&\cg0.42&\cg95.8&\cg0.43\\
&&&&&&&&\\[-0.475cm]
$22.5\cdot10^6$&\cg94.7&\cg0.41&\cg95.6&\cg0.44&\cg94.4&\cg0.42&\cg94.4&\cg0.43\\
&&&&&&&&\\[-0.475cm]
$23.0\cdot10^6$&\cy93.6&\cg0.41&\cy96.4&\cg0.43&\cg95.4&\cg0.41&\cg95.4&\cg0.42\\
&&&&&&&&\\[-0.475cm]
$23.5\cdot10^6$&\cg94.8&\cg0.40&\cg95.1&\cg0.43&\cy96.2&\cg0.41&\cy96.2&\cg0.42\\
&&&&&&&&\\[-0.475cm]
$24.0\cdot10^6$&\cy93.1&\cg0.40&\cg96.0&\cg0.42&\cg94.9&\cg0.41&\cg94.9&\cg0.41\\
&&&&&&&&\\[-0.475cm]
$24.5\cdot10^6$&\cg94.3&\cg0.39&\cy96.7&\cg0.42&\cg95.8&\cg0.40&\cg95.8&\cg0.41\\
&&&&&&&&\\[-0.475cm]
$25.0\cdot10^6$&\cg94.8&\cg0.39&\cg95.5&\cg0.41&\cg94.4&\cg0.40&\cg94.4&\cg0.40\\
&&&&&&&&\\[-0.425cm]
\hline
\end{tabular}
\end{table}

\noindent The $CPr$ and $\tilde{\epsilon}_R$ performance across the $p^*$ range is given in Tables \ref{tab:CIperf1} and \ref{tab:CIperf2}. Shown is the performance of each CI estimator at a selection of sample sizes of interest. The Wald, Wilson and Agresti-Coull methods perform similarly for a given $n\text{-}p^*$ combination, as shown in Table \ref{tab:CIperf1}. The $CPr$ and $\tilde{\epsilon}_R$ values of the Clopper-Pearson slightly exceed the desired limits, but overall, the performance is quite reasonable. 
\pagebreak

\begin{table}[!h]
\footnotesize
\centering
\caption{$95\%$ CI performance comparison - $n$}
\label{tab:CIperf1}
\begin{tabular}{cccccccccc} 
\\[-0.4cm]
\hline\\[-0.4cm]
\multirow{2}{*}{${p^*}$}&\multirow{2}{*}{${n}$}&\multicolumn{2}{c}{{W}}&\multicolumn{2}{c}{{CP}}&\multicolumn{2}{c}{{WS}}&\multicolumn{2}{c}{{AC}}\\
&&{${CPr}$}&$\tilde{\epsilon}_R$&{${CPr}$}&$\tilde{\epsilon}_R$&{${CPr}$}&$\tilde{\epsilon}_R$&{${CPr}$}&$\tilde{\epsilon}_R$\\
\hline\\[-0.41cm]
\multirow{2}{*}{$10^{-1}$}
&$n_A$: $1.4\cdot10^{2}$&\cy93.3&\cg0.49&\cy96.8&\cy0.53&\cg95.2&\cg0.50&\cg95.2&\cy0.51\\ 
&&&&&&&&& \\[-0.475cm]
&$n_B$: $1.6\cdot10^{2}$&\cg94.2&\cg0.47&\cg95.6&\cg0.50&\cg94.1&\cg0.47&\cg95.4&\cg0.48\\
\hline\\[-0.355cm]
&&&&&&&&& \\[-0.475cm]
\multirow{2}{*}{$10^{-2}$}
&$n_A$: $1.6\cdot10^{3}$&\cg94.0&\cg0.49&\co97.1&\cy0.52&\cg95.9&\cg0.50&\cg95.9&\cy0.52\\ 
&&&&&&&&& \\[-0.475cm]
&$n_B$: $1.8\cdot10^{3}$&\cg94.1&\cg0.45&\cg95.4&\cg0.48&\cg95.4&\cg0.46&\cg95.4&\cg0.47\\ 
\hline\\[-0.355cm]
&&&&&&&&& \\[-0.475cm]
\multirow{2}{*}{$10^{-3}$}
&$n_A$: $1.6\cdot10^{4}$&\cg94.0&\cg0.49&\cy97.0&\cy0.53&\cg95.7&\cg0.50&\cg95.7&\cy0.52\\ 
&&&&&&&&& \\[-0.475cm]
&$n_B$: $1.7\cdot10^{4}$&\cg94.7&\cg0.47&\cg95.9&\cg0.50&\cg94.7&\cg0.48&\cg95.9&\cg0.49\\ 
\hline\\[-0.355cm]
&&&&&&&&& \\[-0.475cm]
\multirow{2}{*}{$10^{-4}$}
&$n_A$: $1.6\cdot10^{5}$&\cg94.0&\cg0.49&\cy96.9&\cy0.53&\cg95.7&\cy0.51&\cg95.7&\cy0.52\\ 
&&&&&&&&& \\[-0.475cm]
&$n_B$: $1.7\cdot10^{5}$&\cg94.7&\cg0.47&\cg95.9&\cg0.50&\cg94.7&\cg0.48&\cg95.8&\cg0.49\\ 
\hline\\[-0.355cm]
&&&&&&&&& \\[-0.475cm]
\multirow{2}{*}{$10^{-5}$}
&$n_A$: $1.6\cdot10^{6}$&\cg94.0&\cg0.49&\cy96.9&\cy0.53&\cg95.7&\cy0.51&\cg95.7&\cy0.52\\
&&&&&&&&& \\[-0.475cm]		
&$n_B$: $1.7\cdot10^{6}$&\cg94.7&\cg0.47&\cg95.9&\cg0.50&\cg94.7&\cg0.48&\cg95.9&\cg0.49\\ 
\hline\\[-0.355cm]
&&&&&&&&& \\[-0.475cm]
\multirow{2}{*}{$10^{-6}$}
&$n_A$: $1.6\cdot10^{7}$&\cg94.0&\cg0.49&\cy96.9&\cy0.53&\cg95.7&\cy0.51&\cg95.7&\cy0.52\\
&&&&&&&&& \\[-0.475cm]		
&$n_B$: $1.7\cdot10^{7}$&\cg94.7&\cg0.47&\cg95.9&\cg0.50&\cg94.7&\cg0.48&\cg95.8&\cg0.49\\ 
&&&&&&&&& \\[-0.425cm]
\hline
\multicolumn{10}{l}{\scriptsize $n_A$: First $n$ where at least one interval satisfies $CPr\in95\pm1\%$ and $\tilde{\epsilon}_R\le0.5$}\\[-0.1cm]
\multicolumn{10}{l}{\scriptsize $n_B$: First $n$ where all intervals satisfy $CPr\in95\pm1\%$ and $\tilde{\epsilon}_R\le0.5$}\\[-0.1cm]
\end{tabular}
\end{table}

Table \ref{tab:CIperf2} gives the sample sizes required to maintain a desired level of $CPr$ and $\tilde{\epsilon}_R$. Three performance schemes were investigated: (i) $\hspace{0.1cm}CPr\in95\pm3\%, \tilde{\epsilon}_R\le1$, (ii) $\hspace{0.1cm}CPr\in95\pm2\%, \tilde{\epsilon}_R\le0.75$ and (iii)$\hspace{0.1cm}CPr=95\pm1\%, \tilde{\epsilon}_R\le0.5$. For each scheme, the Wilson interval required the smallest sample size to achieve (and maintain) the desired performance, thus providing further evidence of its overall superiority among the four estimators.
\pagebreak

\begin{table}[!h]
\footnotesize
\centering
\caption{$95\%$ CI sample size comparison }
\label{tab:CIperf2}
\begin{tabular}{cccccc} 
\\[-0.4cm]
\hline\\[-0.4cm]
${p^*}$&${n}$&W&CP&WS&AC\\
\hline\\[-0.41cm]
\multirow{3}{*}{$10^{-1}$}
&$n_C$&\co$1.5\cdot10^{2}$  &\co$5.4\cdot10^{1}$  &\co$\bm{3.6\cdot10^{1}}$ &\co$4.2\cdot10^{1}$  \\
&$n_D$&\cy$2.9\cdot10^{2}$  &\cy$1.9\cdot10^{2}$  &\cy$\bm{6.2\cdot10^{1}}$  &\cy$1.0\cdot10^{2}$  \\
&$n_E$&\cg$6.5\cdot10^{2}$  &\cg$1.2\cdot10^{3}$  &\cg$\bm{3.5\cdot10^{2}}$ &\cg$\bm{3.5\cdot10^{2}}$  \\
&&&&&\\[-0.45cm]
\hline\\[-0.41cm]
\multirow{3}{*}{$10^{-2}$}
&$n_C$&\co$1.5\cdot10^{3}$  &\co$7.1\cdot10^{2}$  &\co$\bm{4.4\cdot10^{2}}$  &\co$5.0\cdot10^{2}$\\
&$n_D$&\cy$2.8\cdot10^{3}$  &\cy$2.0\cdot10^{3}$  &\cy$\bm{7.4\cdot10^{2}}$  &\cy$1.2\cdot10^{3}$\\
&$n_E$&\cg$8.0\cdot10^{3}$  &\cg$1.0\cdot10^{4}$  &\cg$\bm{2.6\cdot10^{3}}$  &\cg$4.1\cdot10^{3}$\\
&&&&&\\[-0.45cm]
\hline\\[-0.41cm]
\multirow{3}{*}{$10^{-3}$}
&$n_C$&\co$1.8\cdot10^{4}$&\co$7.0\cdot10^{3}$&\co$\bm{4.5\cdot10^{3}}$&\co$5.1\cdot10^{3}$\\
&$n_D$&\cy$2.8\cdot10^{4}$&\cy$1.7\cdot10^{4}$&\cy$\bm{7.5\cdot10^{3}}$&\cy$1.2\cdot10^{4}$\\
&$n_E$&\cg$7.4\cdot10^{4}$&\cg$1.0\cdot10^{5}$&\cg$\bm{2.6\cdot10^{4}}$&\cg$3.7\cdot10^{4}$\\
&&&&&\\[-0.45cm]
\hline\\[-0.41cm]
\multirow{3}{*}{$10^{-4}$}
&$n_C$&\co$1.8\cdot10^{5}$&\co$7.0\cdot10^{4}$&\co$\bm{4.5\cdot10^{4}}$&\co$5.1\cdot10^{4}$\\
&$n_D$&\cy$2.8\cdot10^{5}$&\cy$1.6\cdot10^{5}$&\cy$\bm{7.5\cdot10^{4}}$&\cy$1.2\cdot10^{5}$\\
&$n_E$&\cg$7.4\cdot10^{5}$&\cg$1.0\cdot10^{6}$&\cg$\bm{2.9\cdot10^{5}}$&\cg$4.7\cdot10^{5}$\\
&&&&&\\[-0.45cm]
\hline\\[-0.41cm]
\multirow{3}{*}{$10^{-5}$}
&$n_C$&\co$1.8\cdot10^{6}$&\co$7.0\cdot10^{5}$&\co$\bm{4.5\cdot10^{5}}$&\co$5.1\cdot10^{5}$\\
&$n_D$&\cy$2.8\cdot10^{6}$&\cy$1.7\cdot10^{6}$&\cy$\bm{7.5\cdot10^{5}}$&\cy$1.2\cdot10^{6}$\\
&$n_E$&\cg$7.4\cdot10^{6}$&\cg$1.0\cdot10^{7}$&\cg$\bm{2.9\cdot10^{6}}$&\cg$4.7\cdot10^{6}$\\
&&&&&\\[-0.45cm]
\hline\\[-0.41cm]
\multirow{3}{*}{$10^{-6}$}
&$n_C$&\co$1.8\cdot10^{7}$&\co$7.0\cdot10^{6}$&\co$\bm{4.5\cdot10^{6}}$&\co$5.1\cdot10^{6}$\\
&$n_D$&\cy$2.8\cdot10^{7}$&\cy$1.7\cdot10^{7}$&\cy$\bm{7.5\cdot10^{6}}$&\cy$1.2\cdot10^{7}$\\
&$n_E$&\cg$7.4\cdot10^{7}$&\cg$1.0\cdot10^{8}$&\cg$\bm{2.9\cdot10^{7}}$&\cg$4.7\cdot10^{7}$\\

\hline\\[-0.41cm]
\multicolumn{6}{l}{\scriptsize $n_C$: $n_1$ where $CPr\in95\pm3\%$ and $\tilde{\epsilon}_R\le1 \hspace{0.1cm}\forall \hspace{0.1cm}n \ge n_1$}\\[-0.1cm]
\multicolumn{6}{l}{\scriptsize $n_D$: $n_1$ where $CPr\in95\pm2\%$ and $\tilde{\epsilon}_R\le0.75 \hspace{0.1cm}\forall \hspace{0.1cm}n \ge n_1$}\\[-0.1cm]
\multicolumn{6}{l}{\scriptsize $n_E$: $n_1$ where $CPr\in95\pm1\%$ and $\tilde{\epsilon}_R\le0.5 \hspace{0.1cm}\forall \hspace{0.1cm}n \ge n_1$}\\[-0.1cm]
\multicolumn{6}{l}{\scriptsize Smallest sample size for each scheme shown in {bold}}
\end{tabular}
\end{table}

Table \ref{tab:CIperf3} shows the $\tilde{\epsilon}_R$ values pertaining to the sample sizes displayed in Table \ref{tab:CIperf2} ($\epsilon_R$ values were found to be very similar to the given $\tilde{\epsilon}_R$ values). It can be seen that in relation to a $95\%$ CI, to ensure that the coverage remains within $\pm2\%$, the Wald interval requires $\tilde{\epsilon}_R\le 0.37$, whereas the Wilson interval requires $\tilde{\epsilon}_R = 0.75$. The $\tilde{\epsilon}_R$ values corresponding to maintaining the coverage within $95\pm1\%$ (green table cells) are in close agreement with our recommendation to use $\tilde{\epsilon}_R \in [0.1, 0.5]$. 

\begin{table}[!h]
\footnotesize
\centering
\caption{$\tilde{\epsilon}_R$ values corresponding to maintained coverage}
\label{tab:CIperf3}
\begin{tabular}{ccccc} 
\\[-0.4cm]
\hline\\[-0.4cm]
${CPr}$&W&CP&WS&AC\\
\hline\\[-0.41cm]
$95\pm3\%$&\co 0.46 - 0.50 &\co 0.80 - 0.87 &\co 0.98 - 1.00 &\co 0.97 - 1.00 \\
$95\pm2\%$&\cy 0.34 - 0.37 &\cy 0.45 - 0.52 &\cy 0.75 - 0.75 &\cy 0.60 - 0.61 \\
$95\pm1\%$&\cg 0.22 - 0.23 &\cg 0.17 - 0.20 &\cg 0.31 - 0.39 &\cg 0.29 - 0.33 \\
&&&&\\[-0.42cm]
\hline\\[-0.42cm]
\multicolumn{5}{l}{\scriptsize $\tilde{\epsilon}_R$ values observed across the $p^*$ range displayed as minimum - maximum}
\end{tabular}
\end{table}


\section{{Estimating a Rare Event with a Small Sample Size}}
\label{sec:Smalln}

It is clear from the above results, that as expected \textit{a priori}, quite large sample sizes are required to accurately estimate rare-event probabilities. Achieving accuracy on the order of magnitude for a small $p$ is usually most relevant in large populations, where it will also be possible to collect large samples. For example, a quality engineer may have little problem in obtaining high-throughput process data of order $n=10^6$ or greater, and, in this large-scale production setting, it will be critical to know whether the defect rate is, say, one in one thousand, or one in ten thousand. In Section \ref{sec:Data}, we assess three data-rich scenarios from the literature, i.e., cases that involve estimating a small proportion whilst utilizing large samples.

Notwithstanding the fact that accurate estimation of a small $p$ is most important in large populations, an analyst may be faced with the challenge of estimating a small $p$ with a limited sample size. We touched on this problem in Section \ref{sec:zeroevent}, but now consider CI performance in more detail using the approach of Section \ref{sec:CITable}.

Assume that the true proportion is of order $10^{-2}$. As we have seen previously in Table \ref{tab:CIperf2}, sample sizes of order $n=10^{3}$ will be needed to accurately estimate $p$. However, here, we assume that the analyst is dealing with a hard-to-reach population where $n \le 100$; the performance of the four intervals is displayed in Table \ref{tab:p10-2_smalln}.

\begin{table}[!h]
\footnotesize
\centering
\caption{$95\%$ CI performance - ${p^*}=p=10^{-2}$, small $n$}
\label{tab:p10-2_smalln}
\begin{tabular}{ccccccccc} 
\\[-0.4cm]
\hline\\[-0.4cm]
\multirow{2}{*}{${n}$}&\multicolumn{2}{c}{{W}}&\multicolumn{2}{c}{{CP}}&\multicolumn{2}{c}{{WS}}&\multicolumn{2}{c}{{AC}}\\
&{${CPr}$}&${\tilde{\epsilon}_R}$&{${CPr}$}&${\tilde{\epsilon}_R}$&{${CPr}$}&${\tilde{\epsilon}_R}$&{${CPr}$}&${\tilde{\epsilon}_R}$\\
\hline\\[-0.41cm]

20&\cre18.2&\cre1.80&\cre98.3&\cre9.19&\cre98.3&\cre8.70&\cre98.3&\cre11.36\\
&&&&&&&&\\[-0.475cm]

40&\cre33.1&\cre1.73&\cre99.3&\cre5.22&\cy93.9&\cre5.08&\cre99.3&\cre6.57\\
&&&&&&&&\\[-0.475cm]

60&\cre45.2&\cre1.65&\co97.8&\cre3.80&\co97.8&\cre3.72&\co97.8&\cre4.72\\
&&&&&&&&\\[-0.475cm]

80&\cre55.2&\cre1.57&\cre99.1&\cre3.07&\cg95.3&\cre3.00&\cre99.1&\cre3.74\\
&&&&&&&&\\[-0.475cm]

100&\cre63.3&\cre1.50&\cre98.2&\cre2.62&\co92.1&\cre2.55&\cre98.2&\cre3.13\\
&&&&&&&&\\[-0.43cm]
\hline
\end{tabular}
\end{table}

It is clear that all four intervals perform quite poorly in this scenario both in terms of coverage and relative margin of error. The coverage of the Wilson interval is notably better than the others, and is reasonable for some sample sizes, albeit is still somewhat erratic. This interval does achieve excellent coverage for $n = 80$ for example, but the relative margin of error is $\tilde \epsilon_R \approx 3$, i.e., the margin of error of $0.03$ is much larger than $p = 0.01$. If the analyst only requires a rough estimate of $p$, for example, to answer the question of whether or not it is less than 0.1, then such a large margin of error will be acceptable. On the other hand, if the aim to is accurately estimate the order of magnitude of $p$, this will not be achievable for such small sample sizes (and, clearly, performance will degrade further for even smaller $p$). This again highlights the importance of considering relative margin of error in the small $p$ setting, and our suggestion is to use $\tilde \epsilon_R \in [0.1, 0.5]$.

An anonymous reviewer advised us of two modern CI estimation approaches: an asymptotic method based on generalized fiducial inference (GFI) \citep{Hannig:2009}, and a recently-developed exact method known as the ``repro samples'' method \citep{Xie:2022}. The GFI method is an extension of the fiducial argument proposed by \cite{Fisher:1930}, and the repro method is a simulation-based method that provides a finite sample CI coverage guarantee, which is particularly useful in small samples. We have tested both of these more modern methods (see Appendix \ref{app:C}), and have found that they provide reasonable coverage (starting from a conservative position akin to the exact Clopper-Pearson method). However, when $p$ and $n$ are small, the methods experience the same issues as the classical methods we have considered; in particular, the relative margin of error is too large to be used in settings where the order of magnitude of a small $p$ is of interest.  (It is noteworthy, however, that the GFI and repro sampling methods are general inference procedures that provide good finite-sample performance in a wide range of problems beyond proportion estimation.)  Ultimately, all of our work points to the fact that large samples are required in this small-$p$ setting, and we have provided guidelines in Section \ref{sec:CITable}.


\section{Case Studies}
\label{sec:Data}

We now consider the use of the relative margin of error in the estimation of small/rare-event proportions using data from the literature. More specifically, we consider: (i) a study on the prevalence of ADHD prescriptions in adolescents, (ii) a clinical trial relating to COVID-19 vaccine efficacy, and (iii) accident data from commercial jet aircraft records.

Using the values of $n$ and $\widehat{p}$ reported in each of the aforementioned studies, we evaluate the validity of a $95\%$ Wald CI in terms of the interval's relative margin of error. We discuss the Wald interval as it is the most commonly used interval estimator, and for each of these case studies, it produces similar results to the Clopper-Pearson, Wilson and Agresti-Coull intervals. We also refer to our sample size calculations/CI performance analyses to assess the suitability of the sample size in terms of achieving the desired coverage.


\subsection{Assessing Prevalence of ADHD Medication}

The first study we consider is a study conducted by \cite{Sawyer:2017} to assess the prevalence of stimulant and antidepressant medication in Australian children and adolescents with symptoms of ADHD (Attention-Deficit/Hyperactivity Disorder) and major depressive disorder (MDD). A nationally representative sample of $n=$ 6,310 children between the age of $4$ and $17$ was obtained, which found that $13.7\%$ of those with symptoms meeting the criteria of ADHD had used stimulant medications.

For a sample size of $n=$ 6,310, and an estimated proportion of magnitude $\widehat{p}=0.137$, the $95\%$ Wald CI is given as: $[0.129, 0.145]$, with a realized relative margin of error of $\widehat{\epsilon_R}=\widehat{\epsilon}/\widehat{p}\approx0.062$.  Using the Delta method (see Appendix \ref{app:D}), a $95\%$ CI for $\epsilon_R$ is given as $[0.060, 0.064]$. (For each case study, a CI for $\tilde{\epsilon}_R$ was obtained using Monte Carlo simulation in conjunction with the Delta method, and each was found to be in agreement with the corresponding CI estimate for $\epsilon_R$.) For this study, the $\epsilon_R$ CI values fall outside of our recommended range of $\epsilon_R \in [0.1,0.5]$ meaning that the interval is somewhat narrower than what we recommend, i.e., one could achieve an acceptable result with fewer observations. Indeed, Table \ref{tab:data} displays the sample sizes for a selection of $\widehat{\epsilon_R}$ values in this range, and, note, for example, that $\widehat{\epsilon_R}=0.4$ leads to a sample size approximately forty times smaller than the sample size used in the study.
 
\begin{table}[!h]
\footnotesize
\centering
\caption{$95\%$ Wald CI - $\widehat{\epsilon_R}$ comparison}
\label{tab:data}
\begin{tabular}{crccc} 
\\[-0.4cm]
\hline\\[-0.4cm]
$\widehat{p}$ & \multicolumn{1}{c}{$n$}& $\widehat{\epsilon}$ & $\widehat{\epsilon_R}$ & $\widehat{p}\pm\widehat{\epsilon}$\\
\hline\\[-0.39cm]
\multirow{5}{*}{$0.137$}
&6310&0.008&0.06&$[0.129,0.145]$ \\
&&&&\\[-0.425cm]
&605&0.027&0.20&$[0.110,0.164]$ \\
&&&&\\[-0.425cm]
&269&0.041&0.30&$[0.096,0.178]$ \\
&&&&\\[-0.425cm]
&152&0.055 &0.40&$[0.082,0.192]$\\
&&&&\\[-0.425cm]
&97&0.068&0.50&$[0.069,0.205]$ \\
&&&&\\[-0.425cm]
\hline\\[-0.4cm]
\end{tabular}
\end{table}

It can also be seen from Table \ref{tab:data} that a relative margin of error of $\widehat{\epsilon_R}=0.2$ corresponds to a CI which is similar to that computed at $\widehat{\epsilon_R}=0.06$, but uses a sample size that is approximately ten times smaller. Had the order of magnitude of $p$ been known in advance (e.g., if it was known that $p\approx1/10$, rather than $p\approx1/100$), then a smaller sample size would have sufficed. When $p$ is of the order $10^{-2}$, very good coverage is achieved for $n=$ 1,800 (see Table \ref{tab:CIperf1}), hence the study sample size of $n=$ 6,310 is more than adequate for the scenarios where $p=10^{-1}$ and $p=10^{-2}$.


\subsection{COVID-19 Vaccine Efficacy}
\label{COVID}

An efficacy trial of the BNT162b2 mRNA COVID-19 vaccine was conducted by \cite{Polack:2020}. In this placebo-controlled, observer-blind trial, 43,548 participants were randomly assigned either the BNT162b2 vaccine or a placebo treatment. Of the 21,720 participants who received the vaccine, there were 8 cases of COVID-19 after the second dose. This leads to an estimated proportion of $\widehat{p}=x/n=$ 8/21,720 $\approx 3.7\cdot10^{-4}$, and, therefore, the $95\%$ Wald CI is given by $[1.1\cdot10^{-4}, 6.2\cdot10^{-4}]$, with $\widehat{\epsilon_R}\approx0.69$, and a $95\%$ CI for $\epsilon_R$ of $[0.421, 0.875]$. As per Section \ref{sec:MoE} we recommend $\epsilon_R \in [0.1, 0.5]$, which is incorporated in the above interval. However, the CI lower bound is very close to our recommended $\epsilon_R$ upper bound, and thus, in our suggested scheme, the computed interval could be questioned with regard to its width.

Aside from having a somewhat large margin of error (relative to $\widehat{p}$), we need to consider the sample size in relation to the order of magnitude of $\widehat{p}$. I.e., we must assess if the sample size is large enough to provide acceptable coverage. For example, if $p$ were $10^{-4}$, Table \ref{tab:CIperf2} indicates that a sample size of the order $10^{5}$ is required, whereas, here, the sample size is of the order $10^{4}$. Indeed, we have calculated that, with $p=3.7\cdot 10^{-4}$ and $n=$ 21,720, the expected coverage is just $89.2\%$. The Clopper-Pearson, Wilson and Agresti-Coull intervals perform better in this $n\text{-}p$ scheme, achieving coverage of $96.9\%, 95.2\%$ and $95.2\%$ (respectively). However, for this $n\text{-}p$ combination, all three intervals have $\widehat{\epsilon_R}>0.72$. Thus, to obtain a CI estimate where the margin of error is more consistent with $\widehat{p}$, and/or to enhance the coverage probability, a larger sample size would be required.


\subsection{Commercial Aircraft Accidents}

A summary of annual commercial jet aircraft flight hours, departures and accidents is provided by \cite{Boeing}. In the year $2021$, there were $21.6$ million aircraft departures with a total of $23$ recorded incidents/accidents. Although \textit{all} aircraft departures and accidents are recorded here, we may still view this as a sample from a larger population of flights that might have taken place (had demand been higher) or indeed for flights in upcoming years (provided that conditions such as aviation regulations and the composition of aircraft fleets remain similar). Therefore, it is still of interest to compute a confidence interval in this scenario, and, irrespective of the specific target population, the data still suffice for the purpose of demonstrating our proposed scheme.

For this data, $\widehat{p}\approx 1.1\cdot10^{-6}$, which gives a $95\%$ Wald CI of $[0.66\cdot10^{-6},1.54\cdot10^{-6}]$, with $\widehat{\epsilon_R}\approx 0.40$ ($95\%$ CI for $\epsilon_R$ of $[0.328, 0.494]$). This relative margin of error is consistent with our recommendation of $\epsilon_R \in [0.1,0.5]$, and as discussed in Section \ref{sec:eps_p_compatibility}, provides a satisfactory estimator. Indeed, following the approach of Table \ref{tab:CIperf1}, we have found that, when $p=1.1\cdot10^{-6}$, all four estimators satisfy $CPr\in95\pm1\%$ and $\tilde{\epsilon}_R\le0.5$ when $n=21.6 \cdot10^{6}$.

In the context of estimating the proportion of aircraft accidents, the analyst has no control over the sample size. That is to say, had a larger sample size been required, one would simply have to wait for more aircraft departures to occur. However, our analysis provides us with reassurance that our computed interval will perform satisfactorily.


\section{Discussion}
\label{sec:Conclusion}

When constructing confidence intervals for small success probabilities it is important that the margin or error, $\epsilon$, be considered relative to the magnitude of the proportion, $p$. Incompatibilities between $\epsilon$ and $p$ can lead to completely unsatisfactory coverage or unnecessarily narrow intervals that require extremely large sample sizes. When dealing with moderate success probabilities, say $p\ge0.2$, this is less important, but in the context of small or rare-event success probabilities, the consideration of $\epsilon$ relative to $p$ is crucial to reduce the possibility of substantial mismatching between $\epsilon$ and $p$. For example, $\epsilon=0.05$ might be considered as valid precision for $p=10^{-1}$, but such a margin of error is not compatible with a proportion of the order $p=10^{-3}$.  

To ensure $\epsilon$ is compatible with the order of magnitude of $p$, we recommend using a relative margin of error scheme, $\epsilon_R$. We suggest restricting the range of values to $\epsilon_R\in[0.1,0.5]$ as higher values lead to imprecision and poor interval coverage, whereas lower values lead to sample sizes that are likely to be impractically large for many studies. Our recommendation of $\epsilon_R\in[0.1,0.5]$ avoids intervals that are impractically wide or restrictively narrow in terms of sample size requirements, and we show that adequate performance is achieved within this range. In contrast to the existing literature, we have highlighted the importance of the relative margin of error, $\epsilon_R$, in conjunction with the empirical coverage, when assessing CI performance in the small-$p$ setting. When both criterion are considered simultaneously the Wald, Clopper-Pearson (exact), Wilson and Agresti-Coull intervals perform similarly in many cases. In general, all four intervals fail to satisfy both criteria when the sample size is small, with improved performance at larger sample sizes as expected. For example, for a $95\%$ confidence interval when ${p}=10^{-1}$, none of the methods produce a satisfactory interval for $10 \le n \le 140$. Each interval achieves the nominal coverage of $95\%$ at some (albeit not all) sample sizes in this range, but in each case the desired limit of $\epsilon_R \le0.5$ is exceeded. Once the sample size is increased ($n\ge150$), and the $\epsilon_R$ requirement is satisfied, all four intervals perform well in terms of coverage. 

The coverage probabilities of the Wald and Clopper-Pearson intervals for small $n$ are generally poor, particularly in comparison to the Wilson and Agresti-Coull intervals. However, the considerable difference in coverage in such situations is rendered immaterial once the (we believe reasonable) requirement that $\epsilon_R\le0.5$ is considered. When satisfactory performance is defined as achieving a desired $CPr$ \textit{and} $\epsilon_R$, the performance across these commonly-used intervals is much more comparable, particularly if one considers empirical coverage in the range $(1-\alpha)100\pm2\%$. In this relative margin of error framework the criticisms of inadequate coverage for the Wald interval, and excessive conservatism for the Clopper-Pearson interval, are somewhat alleviated, and all four intervals perform quite similarly. Although there are performance similarities, the Wilson and Agresti-Coull intervals are generally superior to the intervals of Wald and Clopper-Pearson. The Wilson and Agresti-Coull intervals achieve similar $CPr$ and $\epsilon_R$ values for given $n\text{-}p\text{-}\alpha$ combinations, however the Wilson interval is narrower and achieves favourable performance at lower sample sizes. 

When the success probability is small, failure to consider the margin of error relative to the order of magnitude of the estimated proportion can result in poor coverage, and/or intervals which are unnecessarily narrow or excessively wide. As shown in the case studies presented in Section \ref{sec:Data}, the relative margin of error criterion provides a simple and effective assessment of the validity of an estimated interval in terms of its width/margin of error. For example, we have shown that all of the interval estimators considered in this paper performed poorly for the COVID-19 study (Section \ref{COVID}) in terms of the relative margin or error, meaning that the confidence intervals were all impractically wide --- and the Wald interval also had notably poor coverage. It is important to ensure that the interval precision is compatible with the order of magnitude of $p$. The relative margin of error serves as a useful evaluation criterion in this regard, and as such, we suggest that it should be considered when planning statistical studies.


\bibliography{References}

\pagebreak
\section*{Appendix}
\appendix


\section{Sample Size Formulae}
\label{app:A}

\noindent \underline{\textit{Clopper-Pearson Interval}}\\[0.01cm]

\noindent Letting $n_{LB}$ be the minimum $n$ satisfying: $p^*-\text{Beta}\left({\alpha}/{2}; np^*,n(1-p^*)+1\right) \le \epsilon$ and letting $n_{UB}$ be the minimum $n$ satisfying: $\text{Beta}\left(1-{\alpha}/{2};np^*+1,n(1-p^*)\right)-p^*\le\epsilon$, the sample size is given by:
\be n =\left\lceil\text{max}\{n_{LB},n_{UB}\}\right\rceil\ee

\noindent \underline{\textit{Wilson Interval}}

\begin{align*} \epsilon&=\dfrac{z_{\alpha/2}\sqrt{{p(1-p)}/{n}+{z_{\alpha/2}^2}/{(4n^2)}}}{1+{z_{\alpha/2}^2}/{n}}\implies\\ n&=\left\lceil\text{max}\left\{\dfrac{z_{\alpha/2}^2}{2\epsilon^2}\left({p^*(1-p^*)}-{2\epsilon^2}\pm\sqrt{\epsilon^2(1-4p^*(1-p^*))+(p^*(1-p^*))^2}\right)\right\}\right\rceil\end{align*}

\vspace{0.5cm}
\noindent \underline{\textit{Agresti-Coull Interval}}\\[0.1cm]

\noindent Letting $\widetilde{p}=({np+z_{\alpha/2}^2/2})/({n+z_{\alpha/2}^2})$ and $\widetilde{n}=n+z_{\alpha/2}^2$:

\begin{align*} \epsilon=z_{\alpha/2}\sqrt{\dfrac{\widetilde{p}(1-\widetilde{p})}{\widetilde{n}}} \implies 
n=\left\lceil\text{max}\left\{-\dfrac{1}{3a}\left(b+\xi^kC+\dfrac{\Delta_0}{\xi^kC}\right)\right\}\right\rceil;  k \in \{0,1,2\}\end{align*}

\noindent where\\
$a=4\epsilon^2$\\
$b=4z_{\alpha/2}^2(3\epsilon^2-p^*(1-p^*))$\\
$\xi = (-1+\sqrt{-3})/2$\\
$C= \sqrt[3]{\left(\Delta_1\pm\sqrt{\Delta_1^2-4\Delta_0^3}\right)/2}$\\
$\Delta_0=16z_{\alpha/2}^4(3\epsilon^2-p^*(1-p^*))^2-24\epsilon^2z_{\alpha/2}^4(6\epsilon^2-1)$\\
$\Delta_1=128z_{\alpha/2}^6(3\epsilon^2-p^*(1-p^*))^3+432\epsilon^4z_{\alpha/2}^6(4\epsilon^2-1)-288\epsilon^2z_{\alpha/2}^6(3\epsilon^2-p^*(1-p^*))(6\epsilon^2-1)$


\section{Fixed Margin of Error Schemes}
\label{app:B}

Table \ref{tab:fixedmoe2} shows that for the fixed margin of error scheme: $\epsilon=4\cdot10^{-4}, p^*=0.5$, all four intervals produce coverage close to the nominal value. For the remaining three margin of error schemes the Wald interval produces very poor coverage in comparison to the other intervals. For example, in \textit{Scheme 1} ($\epsilon=4\cdot10^{-2},p^*=p$), the Wald coverage for $p=10^{-2}$ is just $21.4\%$, whereas the coverage of the Clopper-Pearson, Wilson and Agresti-Coull intervals is $97.7\%$. 

Whilst the Clopper-Pearson, Wilson and Agresti-Coull intervals achieve better coverage, all four intervals fail to produce satisfactory coverage if the margin of error scheme is not compatible with the magnitude of $p$. For example, in \textit{Scheme 3} ($\epsilon=4\cdot10^{-4}, p^*=p$) the Wald interval produces a coverage of just $0.2\%$ for $p=10^{-5}$. However, the Clopper-Pearson, Wilson and Agresti-Coull intervals achieve coverage of $99.8\%$, $99.8\%$ and $100\%$ respectively. Whilst these coverage probabilities are much closer to the desired $95\%$ coverage than the Wald's $0.2\%$, they are too far from the nominal value to be of practical use.\\

\begin{table}[!h]
\footnotesize
\centering
\caption{Coverage comparison - fixed $\epsilon$}
\label{tab:fixedmoe2}
\begin{tabular}{cccccc} 
\\[-0.6cm]
\hline\\[-0.35cm]
MoE Scheme&{${p}$}&W&CP&WS&AC\\
\hline\\[-0.35cm]
\multirow{5}*{1. $\epsilon=4\cdot10^{-2},\hspace{0.1cm} p^*=p$}&$10^{-1}$&93.8&96.9&94.7&95.9\\
&$10^{-2}$&21.4&97.6&97.6&97.6\\
&&&&&\\[-0.4cm]
&$10^{-3}$&00.3&99.7&99.7&99.7\\
&&&&&\\[-0.4cm]
&$10^{-4}$&00.0&100&100&100\\
&&&&&\\[-0.4cm]
&$10^{-5}$&00.0&100&100&100\\
&&&&&\\[-0.4cm]
&$10^{-6}$&00.0&100&100&100\\
&&&&&\\[-0.4cm]
\hline\\[-0.4cm]
\multirow{5}*{2. $\epsilon=4\cdot10^{-2},\hspace{0.1cm} p^*=0.5$}&$10^{-1}$&94.5&95.9&95.2&95.2\\
&$10^{-2}$&93.1&96.3&94.1&97.8\\
&&&&&\\[-0.4cm]
&$10^{-3}$&45.2&97.7&97.7&99.7\\
&&&&&\\[-0.4cm]
&$10^{-4}$&05.8&99.8&94.2&100\\
&&&&&\\[-0.4cm]
&$10^{-5}$&00.6&99.4&99.4&100\\
&&&&&\\[-0.4cm]
&$10^{-6}$&00.1&99.9&99.4&100\\
&&&&&\\[-0.4cm]
\hline\\[-0.4cm]
\multirow{5}*{3. $\epsilon=4\cdot10^{-4},\hspace{0.1cm} p^*=p$}&$10^{-1}$&95.0&95.0&95.0&95.0\\
&$10^{-2}$&95.0&95.1&95.0&95.1\\
&&&&&\\[-0.4cm]
&$10^{-3}$&93.1&96.0&94.9&94.9\\
&&&&&\\[-0.4cm]
&$10^{-4}$&21.3&97.5&97.5&99.8\\
&&&&&\\[-0.4cm]
&$10^{-5}$&00.2&99.8&99.8&100\\
&&&&&\\[-0.4cm]
&$10^{-6}$&00.0&100&100&100\\
&&&&&\\[-0.4cm]
\hline\\[-0.4cm]
\multirow{5}*{4. $\epsilon=4\cdot10^{-4},\hspace{0.1cm} p^*=0.5$}&$10^{-1}$&95.0&95.0&95.0&95.0\\
&$10^{-2}$&95.0&95.0&95.0&95.0\\
&&&&&\\[-0.4cm]
&$10^{-3}$&95.0&95.1&95.0&95.0\\
&&&&&\\[-0.4cm]
&$10^{-4}$&94.9&95.2&95.0&95.0\\
&&&&&\\[-0.4cm]
&$10^{-5}$&94.9&96.1&95.5&95.5\\
&&&&&\\[-0.4cm]
&$10^{-6}$&93.4&96.3&94.0&97.7\\
&&&&&\\[-0.4cm]
\hline
\end{tabular}
\end{table}

\noindent Referring to Table \ref{tab:varmoe2}, for $\epsilon_R=0.75$ and $p^*=p\le10^{-2}$, the Wald coverage is considerably poorer than the other three intervals. Both the Clopper-Pearson and Agresti-Coull intervals produce reasonable coverage at approximately $97\%$. The Wilson interval performs the best, producing coverage very close to the nominal $95\% \hspace{0.1cm} \forall \hspace{0.1cm}p$.\\

\begin{table}[!h]
\footnotesize
\centering
\caption{Coverage comparison - $\epsilon_R=0.75$}
\label{tab:varmoe2}
\begin{tabular}{ccccc} 
\\[-0.6cm]
\hline\\[-0.4cm]
{${p^*}$}&W&CP&WS&AC\\
\hline\\[-0.36cm]
$10^{-1}$&94.7&97.0&94.6&94.6\\
&&&&\\[-0.4cm]
$10^{-2}$&90.2&97.0&94.9&97.0\\
&&&&\\[-0.4cm]
$10^{-3}$&90.4&96.9&94.6&96.9\\
&&&&\\[-0.4cm]
$10^{-4}$&90.4&96.9&94.6&96.9\\
&&&&\\[-0.4cm]
$10^{-5}$&90.4&96.9&94.6&96.9\\
&&&&\\[-0.4cm]
$10^{-6}$&90.4&96.9&94.6&96.9\\
&&&&\\[-0.4cm]
\hline
\end{tabular}
\end{table}


\section{CI Estimation using GFI and Repro Samples Method}
\label{app:C}

To compare the performance of the Wald, Clopper-Pearson, Wilson and Agresti-Coull methods against the more modern GFI \citep{Hannig:2009} and repro sampling \citep{Xie:2022} methods, we consider $p \in [10^{-1}, 10^{-4}]$, with sample sizes ranging from $n=20$ to $n=10^{4}$. This covers a range of scenarios from the main paper, including larger samples as per Section \ref{sec:CITable} and smaller samples as per Section \ref{sec:Smalln}. As these more modern methods are simulation-based, this makes exact performance calculations more challenging (as were done in the main paper). Therefore, here, we have carried out a simulation study with a large number of replicates (5000). The results are presented in Table \ref{tab:modern}.

We can see from Table \ref{tab:modern}, that, in terms of coverage, both of these methods converage towards the nominal $95\%$ level with the sample size (starting from a conservative point akin to the exact Clopper-Pearson interval). Interestingly, in some of the more challenging scenarios presented here, e.g., $p = 10^{-3}$ with $n = 10^3$ and $p = 10^{-4}$ and $n = 10^4$, one or both of these methods provide the best empirical coverage (albeit still being a 2-3 percentage points out from the nominal level in these scenarios). In any case, these methods do not appear to offer dramatic improvements on the other methods considered, and, in particular, the key issue of $\tilde \epsilon_R$ being large for small $n$ is intrinsic to all of the methods. It is important to recognize, however, that the GFI and repro sampling methods are not limited only to proportion estimation, but, rather, are general inferential techniques that extend far beyond this to many other problems.

\begin{table}[!h]
\footnotesize
\centering
\caption{Comparison of classical and modern CI methods}
\label{tab:modern}
\begin{tabular}{cccccccccccccc} 
\\[-0.4cm]
\hline\\[-0.4cm]
\multirow{2}{*}{${p^*}$}&\multirow{2}{*}{${n}$}&\multicolumn{2}{c}{{W}}&\multicolumn{2}{c}{{CP}}&\multicolumn{2}{c}{{WS}}&\multicolumn{2}{c}{{AC}}&\multicolumn{2}{c}{{GFI}}&\multicolumn{2}{c}{{Repro}}\\
&&{${CPr}$}&$\tilde{\epsilon}_R$&{${CPr}$}&$\tilde{\epsilon}_R$&{${CPr}$}&$\tilde{\epsilon}_R$&{${CPr}$}&$\tilde{\epsilon}_R$&{${CPr}$}&$\tilde{\epsilon}_R$&{${CPr}$}&$\tilde{\epsilon}_R$\\
\hline\\[-0.41cm]
\multirow{6}{*}{$10^{-1}$}
&20&\cre 87.6&\cre 1.17&\cre98.9&\cre1.46 &\cg95.7 &\cre1.32 &\cg95.7 &\cre1.46  &\co 97.9&\cre 1.31&\cre 98.2 &\cre 1.42 \\ 
&&&&&&&&& \\[-0.475cm]
&60&\cg 94.1&\cy 0.74 &\co 97.2 &\co 0.82 &\cg95.2 &\co0.76 &\cg95.2 &\co 0.80 &\cg 95.7&\co 0.76&\co 97.1&\co 0.80 \\ 
&&&&&&&&& \\[-0.475cm]
&$10^{2}$&\cy 93.2&\cy 0.58&\cg 95.6&\cy 0.63&\cy 93.6&\cy 0.59&\co 97.2&\cy 0.61 &\cy 96.1&\cy 0.59&\cg 95.8&\cy 0.61\\ 
&&&&&&&&& \\[-0.475cm]
&$10^{3}$&\cg 95.3&\cg 0.19&\cg 95.5&\cg 0.19&\cg 94.9&\cg 0.19&\cg 94.9&\cg 0.19 &\cg 94.8 &\cg 0.19 &\cg 94.6 &\cg 0.18 \\ 
&&&&&&&&& \\[-0.475cm]
&$10^{4}$&\cg 95.1&\cg 0.06&\cg 95.1&\cg 0.06&\cg 94.9&\cg 0.06&\cg 94.9&\cg 0.06 &\cg 94.9&\cg 0.06&\cg 94.9 &\cg 0.06 \\  
\hline\\[-0.355cm]
&&&&&&&&& \\[-0.475cm]

\multirow{6}{*}{$10^{-2}$}
&20&\cre 18.2&\cre 1.80&\cre 98.3&\cre 9.19 &\cre 98.3&\cre 8.70&\cre 98.3&\cre 11.36 &\cre 98.6 &\cre 7.84&\cre 98.5&\cre 9.02 \\ 
&&&&&&&&& \\[-0.475cm]
&60&\cre 45.2 &\cre 1.65&\co 97.8 &\cre 3.80&\co 97.8&\cre 3.72&\co 97.8&\cre 4.72&\co 97.5&\cre 3.26 &\cre 98.3&\cre 3.73 \\ 
&&&&&&&&& \\[-0.475cm]
&$10^{2}$&\cre 63.3&\cre 1.50&\cre 98.2&\cre 2.62&\co 92.1 &\cre 2.55&\cre 98.2&\cre3.13&\cre 98.3 &\cre 2.27 &\cre 98.2&\cre 2.57 \\ 
&&&&&&&&& \\[-0.475cm]
&$10^{3}$&\co 92.7&\cy 0.61&\co 97.6&\cy 0.67&\cy96.4 &\cy 0.64&\cy 96.4&\cy 0.66 &\cg 94.4&\cy 0.62&\cy 96.2&\cy 0.65\\ 
&&&&&&&&& \\[-0.475cm]
&$10^{4}$&\cg 94.6&\cg 0.19&\cg 95.6&\cg 0.20&\cg 95.0&\cg 0.20&\cg 95.0&\cg 0.20 &\cg 94.6&\cg 0.19&\cg 94.9&\cg 0.19 \\ 
\hline\\[-0.355cm]
&&&&&&&&& \\[-0.485cm]

\multirow{6}{*}{$10^{-3}$}
&20&\cre 02.0&\cre 1.90&\cre 98.0 &\cre 85.01&\cre 98.0&\cre 81.22&\cre 100&\cre $>\hspace{-0.1cm}100$ &\cre 98.2&\cre 69.93 &\cre 98.2&\cre 83.91 \\ 
&&&&&&&&& \\[-0.475cm]
&60&\cre 05.8&\cre 1.91&\cre 99.8&\cre 30.69&\cg 94.2&\cre 30.84&\cre 99.8&\cre 42.46&\cre 98.8&\cre 25.05&\cre 99.7&\cre 30.26 \\ 
&&&&&&&&& \\[-0.475cm]
&$10^2$&\cre 09.5 &\cre 1.89&\cre 99.5&\cre 19.00&\cre 90.5&\cre 19.27&\cre 99.5&\cre 26.50 &\cre 99.4&\cre 15.58&\cre 99.5 &\cre 18.82 \\ 
&&&&&&&&& \\[-0.475cm]
&$10^3$&\cre 63.2&\cre 1.51&\cre 98.1&\cre 2.67&\cre 91.9&\cre 2.65&\cre 98.1&\cre 3.29 & \co 97.9&\cre 2.35&\co 97.9&\cre 2.63\\ 
&&&&&&&&& \\[-0.475cm]
&$10^4$&\co 92.6&\cy 0.61&\co 97.5&\cy 0.67&\cy 96.3&\cy 0.64&\cy 96.3&\cy 0.67&\cg 94.4&\cy 0.63&\cy 96.5&\cy 0.66 \\
\hline\\[-0.355cm]
&&&&&&&&& \\[-0.475cm]

\multirow{6}{*}{$10^{-4}$}
&20&\cre 00.2&\cre 1.91&\cre 99.8&\cre $>\hspace{-0.1cm}100$ &\cre 99.8 &\cre $>\hspace{-0.1cm}100$ &\cre 100&\cre $>\hspace{-0.1cm}100$ &\cre 99.8&\cre $>\hspace{-0.1cm}100$&\cre 99.8 & \cre $>\hspace{-0.1cm}100$ \\ 
&&&&&&&&& \\[-0.475cm]
&60&\cre 00.6&\cre 1.94&\cre 99.4&\cre $>\hspace{-0.1cm}100$ &\cre 99.4&\cre $>\hspace{-0.1cm}100$ &\cre 100 &\cre $>\hspace{-0.1cm}100$ &\cre 99.4&\cre $>\hspace{-0.1cm}100$&\cre 99.4 &\cre $>\hspace{-0.1cm}100$ \\  
&&&&&&&&& \\[-0.475cm]
&$10^2$&\cre 01.0 &\cre 1.94 &\cre 99.0 &\cre $>\hspace{-0.1cm}100$ &\cre 99.0 &\cre $>\hspace{-0.1cm}100$ &\cre 100&\cre $>\hspace{-0.1cm}100$ &\cre 99.3 &\cre $>\hspace{-0.1cm}100$&\cre 98.9&\cre $>\hspace{-0.1cm}100$ \\
&&&&&&&&& \\[-0.475cm]
&$10^3$&\cre 09.5 &\cre 1.90 &\cre 99.5 &\cre 19.33&\cre 90.5&\cre 19.94 &\cre 100 &\cre 27.66 &\cre 99.6&\cre 15.83&\cre 99.4&\cre 19.11 \\ 
&&&&&&&&& \\[-0.475cm]
&$10^4$&\cre 63.2&\cre 1.52&\cre 98.1&\cre 2.68&\cre 91.9 &\cre 2.66&\cre 98.1&\cre 3.29&\cre 98.2&\cre 2.34&\co 97.4&\cre 2.65 \\ 
\hline\\[-0.355cm]
&&&&&&&&& \\[-0.475cm]
\end{tabular}
\end{table}


\section{CI For  \texorpdfstring{$\bm{\epsilon_R}$}{$\epsilon,R$}} 
\label{app:D}

By the Delta method, $f(\widehat{p}) \rightarrow \text{N}(f(p), f'(p)^2\sigma_{\widehat{p}}^2)$ as $n \rightarrow \infty$. Given $f(\widehat{p})=\widehat{\epsilon_R}=\widehat{\epsilon}/\widehat{p}$, and $\sigma_{\widehat{p}}= \sqrt{\widehat{p}(1-\widehat{p})/n}$, a $(1-\alpha)100\%$ Wald CI for $\epsilon_R$ is given by \be \label{eq:CIeps}\dfrac{\tilde{z}}{\sqrt{n}}\sqrt{\dfrac{1-\widehat{p}}{\widehat{p}}} 
\pm z_{\alpha/2}\dfrac{\tilde{z}}{\sqrt{4n\widehat{p}^4}}\sqrt{\dfrac{\widehat{p}}{1-\widehat{p}}}\sqrt{\dfrac{\widehat{p}(1-\widehat{p})}{n}},\ee where $\tilde{z}$ is the $(1-\tilde{\alpha}/2)$ quantile of the standard normal distribution, and $\tilde{\alpha}$ is the significance level pertaining to $\widehat{\epsilon}$.

\end{document}